\documentclass[aps,pra,groupedaddress,twocolumn]{revtex4}

\usepackage{amsmath}
\usepackage{graphicx}	
\usepackage{bm} 
\usepackage{color}
\usepackage{epstopdf}

\begin{document}
\title{Formation of classical crystals of dipolar particles in a
  helical geometry}

\author{J.~K. Pedersen}
\author{D.~V. Fedorov}
\author{A.~S. Jensen}
\author{N.~T. Zinner}
\affiliation{Department of Physics and Astronomy, Aarhus University, DK-8000 Aarhus C, Denmark}

\date{\today}

\begin{abstract}
We consider crystal formation of particles with dipole-dipole
interactions that are confined to move in a one-dimensional helical
geometry with their dipole moments oriented along the symmetry axis of
the confining helix. The stable classical lowest energy configurations
are found to be chain structures for a large range of pitch-to-radius
ratios for relatively low density of dipoles and a moderate total
number of particles.  The classical normal mode spectra support the
chain interpretation both through structure and the distinct
degeneracies depending discretely on the number of dipoles per
revolution.  A larger total number of dipoles leads to a
clusterization where the dipolar chains move closer to each
other. This implies a change in the local density and the emergence of
two length scales, one for the cluster size and one for the
inter-cluster distance along the helix.  Starting from three dipoles
per revolution, this implies a breaking of the initial periodicity to
form a cluster of two chains close together and a third chain removed
from the cluster.  This is driven by the competition between in-chain
and out-of-chain interactions, or alternatively the side-by-side
repulsion and the head-to-tail attraction in the system.  The speed of
sound propagates along the chains. It is independent of the number of
chains although depending on geometry.
\end{abstract}

\maketitle

\section{Introduction}

A highly promising direction in cold atoms phyiscs is the experimental
realization and exploration of homo- and heteronuclear molecules at
low temperatures
\cite{ospelkaus2008,ni2008,deiglmayr2008,lang2008,ospelkaus2010,ni2010,shuman2010,miranda2011,chotia2012}.
The potential applications of such systems are wide-ranging within
quantum simulation, information, computation, and metrology
\cite{baranov2008,lahaye2009,carr2009,baranov2012}.  An important
issue is the head-to-tail attraction that molecules with dipole
moments have which can severely limit experimental timescales due to
strong losses \cite{ospelkaus2010,lushnikov2002}.  The suggested way forward has
been to confine the molecules in low-dimensional setups where the
head-to-tail attraction can be controllably reduced and losses can be
suppressed. An experimental confirmation of this was recently achieved
with a stack of two-dimensional planes \cite{miranda2011} and in a
three-dimensional optical lattice \cite{chotia2012}.

A great advantage of low-dimensional setups is that one can study the
interplay of geometrical restrictions on dynamics in the presence of
long-range interactions from the dipole-dipole forces. This is one of
the main motivations of the present paper. In the context of
condensed-matter systems in low-dimensional systems, this interplay of
range and geometry is a very active area of research (see
Ref.~\cite{giamarchi2004} for an introduction). The geometry we will
consider is that of a helix where we assume that the pitch and radius
can be tuned experimentally.  This can be realized in different
ways. Laguerre-Gaussian beams with non-zero angular momentum are one
possibility (bright 
\cite{macdonald2002,pang2005,bhatta2007,ricardez2010,okulov2012} or dark \cite{arnold2012,beattie2013}).
This method has some less desirable limitations on the maximum length
of the uniform helix and the variational range of the parameters
\cite{reitz2012}. These restrictions can be dealt with by working instead
with light guided by an optical nanofiber.  Here it was recently shown
that atoms can be trapped in the evanescent field surrounding the
fiber waist \cite{sague2008,vetsch2010,dawkins2011}.  A particularly
interesting proposal concerns the generation of the double-helix
potential known from DNA molecules \cite{reitz2012}.

Here we will explore the physics of a single helix setup and assume
that cold molecules are trapped on the helix. A few years ago, the
same geometry was discussed as a possible example of a liquid-gas
transition at zero temperature \cite{law2008}.  Furthermore, studies
of the competition between dipolar particles and the external trapping
potential have been done and linear to zig-zag transition were found
\cite{astrak2008,astrak2009}. This work builds on earlier findings of
similar transitions in the case where the dipolar particles are
replaced by charged ions \cite{fishman2008,chiara2010}. Helical
structures have also been studied in 
spinor Bose-Einstein condensate with dipolar interactions where
they can appear as spin textures \cite{huhta2010}.

\begin{figure}
\centering
\includegraphics[width=\columnwidth]{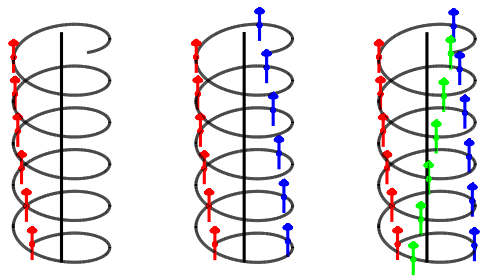}
\caption{Schematic drawings of chain configurations on a helix. Left
  is one chain, middle two chains asymmetrically located along ten
  helix, right is three chains.  To guide the eye the dipoles in each
  chain have the same colours.
 \label{fig:draw2peromgang} }
\end{figure} 

In our work we will assume that the helical trap is rigid and no
movement can occur away from the helix as done in some recent studies
of the charged ions on a helix \cite{schmelcher2011,zampetaki2013}.
This effectively means that we assume a tight trapping potential and
the cost of excitations in the transverse directions will thus be much
larger than the interaction energies.  We illustrate the geometry in
fig.~\ref{fig:draw2peromgang} with different numbers of dipoles
located on the helix.  As we shall see they tend to form structures
organized in chains depending on the density of the particles. The
distance between the chains is not necessarily equidistant because
the attractive and repulsive parts of the interaction strongly depend
on dipole direction and distance.  We are not aware of a previous
discussion of the classical ground state phases in this setup with
dipolar molecules.

In section II we provide energy expressions in a suitable coordinate
system and focus on the simplest configurations as function of
geometric helix parameters and particle number.  In sections III and
IV we discuss stability and normal modes, and the speed of sound in
these 1D systems.  Finally, section V contains a brief summary of
the conclusions in addition to an outlook and perpectives related to experimental
realizations.

\section{Dipoles on a helix}

We consider dipoles moving on the one dimensional geometry of a helix.
An experimental realization of this geometry is possible by applying
counter propagating laser beams to a tapered optical fiber
\cite{reitz2012}, and trapping the dipoles in the evanescent field
around the fiber. This is not strictly one dimensional, but by varying
the laser power one can enter a one dimensional regime.  We first
specify helix parameters, dipole coodinates, potentials and energies,
and second we calculate classically stable configurations.

\subsection{Classical energy}
\begin{figure}
\includegraphics[scale=1.0]{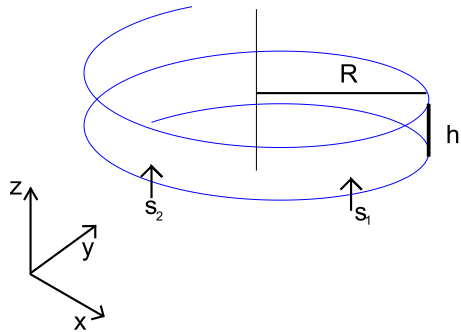}
\caption{Schematic drawing of a helix of radius $R$ and pitch $h$,
  with two dipoles, $s_1$ and $s_2$, directed along the
  $z$-axis.  \label{fig:helix}}
\end{figure}

A helix is described by two parameters, the height $h$ and the radius
$R$. See fig.~\ref{fig:helix} for a schematic drawing of a helix.  The
radius is the distance at which the helix revolves around a fixed
axis, and the height is the vertical distance between two such
revolutions.  The position along the helix can be described by the arc
length, $s$, along the helix or equivalently by the azimuthal angle,
$\phi$, around the z-axis, $ \phi = s/\alpha$, where
$\alpha=\left(R^2+(\frac{h}{2\pi})^2\right)^{1/2}$.  We shall use the
angle $\phi$ where one circle is described by $\phi \in [0,2\pi]$, and
the following by $\phi$-values increasing beyond $2\pi$.  The relation to
Cartesian coordinates are given by the following transformation
\begin{equation}\label{eq:grptrans}
 (x,y,z) = (R\sin \phi,R\cos \phi,h \frac{\phi}{2\pi}) \; .
\end{equation}
On the helix we place $N$ identical dipoles of mass $m$ and dipole
moment $\bm{d}$, all aligned along the z-axis by an external field.
The dipoles are then confined to move only on the helix, but they
still interact through the three dimensional Cartesian space.  The
potential energy, $V$, of two dipoles at position $\bm{r}_i$ and
$\bm{r}_j$ is
\begin{equation}
 V(\bm{r}_i,\bm{r}_j) = \frac{1}{4\pi\epsilon_0}\frac{1}{r^3}\left[\bm{d}
 \cdot\bm{d}-3\left(\bm{d}\cdot\bm{\hat{r}}\right)\left(\bm{d}
 \cdot\bm{\hat{r}}\right)\right],
\end{equation}
where $r=|\bm{r}_i - \bm{r}_j|$ is the distance between the dipoles,
and $\bm{\hat{r}} = (\bm{r}_i - \bm{r}_j)/r $ is the unit vector in the
direction connecting the two dipoles.  Instead of describing the
position of the dipoles by their Cartesian coordinates it turns out to
be easier to use their position along the helix.

Using the transformation in Eq.(\ref{eq:grptrans}) the potential
energy of two dipoles positioned at angles $\phi_i$ and $\phi_j$ on
the helix becomes
\begin{eqnarray}
 & & V(\phi_i,\phi_j)=\frac{d^2}{4\pi\epsilon_0}  \label{eq:potential} 
 \\ \nonumber
  &\times& \frac{2R^2\left[1-
 \cos{(\phi_i-\phi_j)}\right]
 -2 h^2\left((\phi_i-\phi_j)/(2\pi) \right)^2 }
 {\left( 2R^2\left[1-\cos{(\phi_i-\phi_j)}\right]
 + h^2\left((\phi_i-\phi_j)/(2\pi) \right)^2 
 \right)^{5/2}}.
\end{eqnarray} 
This effective potential scales as $R^{-3}$, and depends only on $h/R$
and the size of the relative angle, $|\phi| =
\left|\phi_i-\phi_j\right|$, between the two dipoles.  

\begin{figure}
\centering
\begingroup
  \makeatletter
  \providecommand\color[2][]{%
    \GenericError{(gnuplot) \space\space\space\@spaces}{%
      Package color not loaded in conjunction with
      terminal option `colourtext'%
    }{See the gnuplot documentation for explanation.%
    }{Either use 'blacktext' in gnuplot or load the package
      color.sty in LaTeX.}%
    \renewcommand\color[2][]{}%
  }%
  \providecommand\includegraphics[2][]{%
    \GenericError{(gnuplot) \space\space\space\@spaces}{%
      Package graphicx or graphics not loaded%
    }{See the gnuplot documentation for explanation.%
    }{The gnuplot epslatex terminal needs graphicx.sty or graphics.sty.}%
    \renewcommand\includegraphics[2][]{}%
  }%
  \providecommand\rotatebox[2]{#2}%
  \@ifundefined{ifGPcolor}{%
    \newif\ifGPcolor
    \GPcolortrue
  }{}%
  \@ifundefined{ifGPblacktext}{%
    \newif\ifGPblacktext
    \GPblacktextfalse
  }{}%
  \let\gplgaddtomacro\g@addto@macro
  \gdef\gplbacktext{}%
  \gdef\gplfronttext{}%
  \makeatother
  \ifGPblacktext
    \def\colorrgb#1{}%
    \def\colorgray#1{}%
  \else
    \ifGPcolor
      \def\colorrgb#1{\color[rgb]{#1}}%
      \def\colorgray#1{\color[gray]{#1}}%
      \expandafter\def\csname LTw\endcsname{\color{white}}%
      \expandafter\def\csname LTb\endcsname{\color{black}}%
      \expandafter\def\csname LTa\endcsname{\color{black}}%
      \expandafter\def\csname LT0\endcsname{\color[rgb]{1,0,0}}%
      \expandafter\def\csname LT1\endcsname{\color[rgb]{0,1,0}}%
      \expandafter\def\csname LT2\endcsname{\color[rgb]{0,0,1}}%
      \expandafter\def\csname LT3\endcsname{\color[rgb]{1,0,1}}%
      \expandafter\def\csname LT4\endcsname{\color[rgb]{0,1,1}}%
      \expandafter\def\csname LT5\endcsname{\color[rgb]{1,1,0}}%
      \expandafter\def\csname LT6\endcsname{\color[rgb]{0,0,0}}%
      \expandafter\def\csname LT7\endcsname{\color[rgb]{1,0.3,0}}%
      \expandafter\def\csname LT8\endcsname{\color[rgb]{0.5,0.5,0.5}}%
    \else
      \def\colorrgb#1{\color{black}}%
      \def\colorgray#1{\color[gray]{#1}}%
      \expandafter\def\csname LTw\endcsname{\color{white}}%
      \expandafter\def\csname LTb\endcsname{\color{black}}%
      \expandafter\def\csname LTa\endcsname{\color{black}}%
      \expandafter\def\csname LT0\endcsname{\color{black}}%
      \expandafter\def\csname LT1\endcsname{\color{black}}%
      \expandafter\def\csname LT2\endcsname{\color{black}}%
      \expandafter\def\csname LT3\endcsname{\color{black}}%
      \expandafter\def\csname LT4\endcsname{\color{black}}%
      \expandafter\def\csname LT5\endcsname{\color{black}}%
      \expandafter\def\csname LT6\endcsname{\color{black}}%
      \expandafter\def\csname LT7\endcsname{\color{black}}%
      \expandafter\def\csname LT8\endcsname{\color{black}}%
    \fi
  \fi
  \setlength{\unitlength}{0.0500bp}%
  \begin{picture}(4676.00,4534.00)%
    \gplgaddtomacro\gplbacktext{%
      \csname LTb\endcsname%
      \put(946,704){\makebox(0,0)[r]{\strut{}-2.5}}%
      \put(946,1213){\makebox(0,0)[r]{\strut{}-2}}%
      \put(946,1723){\makebox(0,0)[r]{\strut{}-1.5}}%
      \put(946,2232){\makebox(0,0)[r]{\strut{}-1}}%
      \put(946,2741){\makebox(0,0)[r]{\strut{}-0.5}}%
      \put(946,3250){\makebox(0,0)[r]{\strut{} 0}}%
      \put(946,3760){\makebox(0,0)[r]{\strut{} 0.5}}%
      \put(946,4269){\makebox(0,0)[r]{\strut{} 1}}%
      \put(1538,484){\makebox(0,0){\strut{} 0.5}}%
      \put(2049,484){\makebox(0,0){\strut{} 1}}%
      \put(2559,484){\makebox(0,0){\strut{} 1.5}}%
      \put(3070,484){\makebox(0,0){\strut{} 2}}%
      \put(3581,484){\makebox(0,0){\strut{} 2.5}}%
      \put(4092,484){\makebox(0,0){\strut{} 3}}%
      \put(176,2486){\rotatebox{-270}{\makebox(0,0){\strut{}$V(\phi)$}}}%
      \put(2678,154){\makebox(0,0){\strut{}$\frac{\phi}{2\pi}$}}%
    }%
    \gplgaddtomacro\gplfronttext{%
    }%
    \gplbacktext
    \put(0,0){\includegraphics{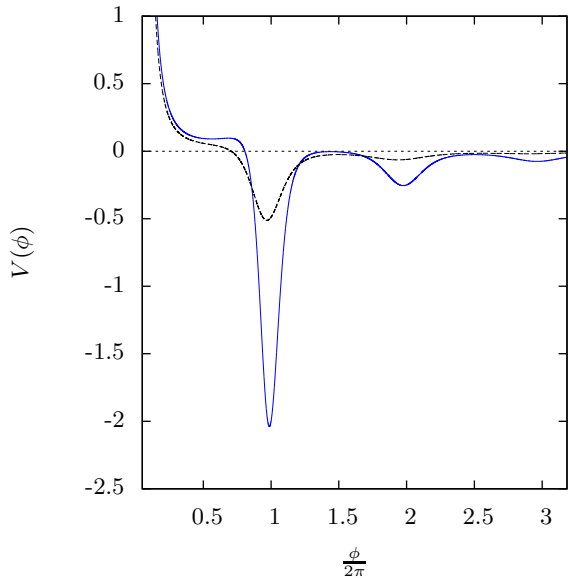}}%
    \gplfronttext
  \end{picture}%
\endgroup
\caption{The reduced potential, $4\pi R^3 \epsilon_0 V /d^2$, of two
  dipoles as a function of the relative angle, $\phi =
  \phi_i-\phi_j$, separating the particles on the helix. The helix
  parameters are chosen to be $h = R$(Blue) and $h = 1.6 R$(black).  \label{fig:potential}}
\end{figure}

We shall from now on use the resulting natural unit, $d^2/(4\pi
\epsilon_0 R^3 )$, for both potentials and energies. The potential
from Eq.(\ref{eq:potential}) in this unit is shown in
fig.~\ref{fig:potential} as function of $\phi$ for intermediate values
of $h$ and $R$.  The repulsion for small $\phi$ when the dipoles
approach each other diverge as $\phi^{-3}$. For large $\phi$ the
attraction vanishes with the same power, $\phi^{-3}$.
 
The strongly damped oscillatory behaviour reveals several minima with
depths decreasing as $\phi^{-3}$, the same power as for
interacting dipoles in three dimensions.  The minima are crudely
located at multipla of $2\pi$, that is the first of them around $6$,
$12$, and $18$ as seen in fig.~\ref{fig:potential}.  Thus, we have an
effective 1D system with a long range dipole-like interaction.

As seen from Eq.(\ref{eq:potential}) the diverging repulsion for small
$\phi$ with $h=R=1$ turns into a diverging $\phi^{-3}$-attraction for
sufficiently large $h/R$-values, that is when $\frac{h}{R}>\pi \sqrt{2}$.  
The two dipoles would then prefer to be in the same point
with infinite energy.  In these cases other short-range forces such as
the van der Waals interaction becomes important.  We want to avoid
this regime where chemistry rules, and hence we limit our
investigations to cases where $\frac{h}{R}<\pi \sqrt{2}$.

The total potential energy, $E$, is obtained by adding the
contributions from all pairs through Eq.(\ref{eq:potential}) for any
set of positions given by the coordinates $\{\phi_k\}$, that is
\begin{equation}
 E(\{\phi_k\})  = \sum_{i<j}^{N}{V(\phi_i,\phi_j)} \;. \label{eq:totalenergy}
\end{equation}
The stable crystal configurations can then be found as minima of the
energy landscape providing equilibrium values for the positions
$\{\phi_k\}$ for any given $N$ as function of $h/R$.

\begin{figure}
\centering
\begingroup
  \makeatletter
  \providecommand\color[2][]{%
    \GenericError{(gnuplot) \space\space\space\@spaces}{%
      Package color not loaded in conjunction with
      terminal option `colourtext'%
    }{See the gnuplot documentation for explanation.%
    }{Either use 'blacktext' in gnuplot or load the package
      color.sty in LaTeX.}%
    \renewcommand\color[2][]{}%
  }%
  \providecommand\includegraphics[2][]{%
    \GenericError{(gnuplot) \space\space\space\@spaces}{%
      Package graphicx or graphics not loaded%
    }{See the gnuplot documentation for explanation.%
    }{The gnuplot epslatex terminal needs graphicx.sty or graphics.sty.}%
    \renewcommand\includegraphics[2][]{}%
  }%
  \providecommand\rotatebox[2]{#2}%
  \@ifundefined{ifGPcolor}{%
    \newif\ifGPcolor
    \GPcolortrue
  }{}%
  \@ifundefined{ifGPblacktext}{%
    \newif\ifGPblacktext
    \GPblacktextfalse
  }{}%
  \let\gplgaddtomacro\g@addto@macro
  \gdef\gplbacktext{}%
  \gdef\gplfronttext{}%
  \makeatother
  \ifGPblacktext
    \def\colorrgb#1{}%
    \def\colorgray#1{}%
  \else
    \ifGPcolor
      \def\colorrgb#1{\color[rgb]{#1}}%
      \def\colorgray#1{\color[gray]{#1}}%
      \expandafter\def\csname LTw\endcsname{\color{white}}%
      \expandafter\def\csname LTb\endcsname{\color{black}}%
      \expandafter\def\csname LTa\endcsname{\color{black}}%
      \expandafter\def\csname LT0\endcsname{\color[rgb]{1,0,0}}%
      \expandafter\def\csname LT1\endcsname{\color[rgb]{0,1,0}}%
      \expandafter\def\csname LT2\endcsname{\color[rgb]{0,0,1}}%
      \expandafter\def\csname LT3\endcsname{\color[rgb]{1,0,1}}%
      \expandafter\def\csname LT4\endcsname{\color[rgb]{0,1,1}}%
      \expandafter\def\csname LT5\endcsname{\color[rgb]{1,1,0}}%
      \expandafter\def\csname LT6\endcsname{\color[rgb]{0,0,0}}%
      \expandafter\def\csname LT7\endcsname{\color[rgb]{1,0.3,0}}%
      \expandafter\def\csname LT8\endcsname{\color[rgb]{0.5,0.5,0.5}}%
    \else
      \def\colorrgb#1{\color{black}}%
      \def\colorgray#1{\color[gray]{#1}}%
      \expandafter\def\csname LTw\endcsname{\color{white}}%
      \expandafter\def\csname LTb\endcsname{\color{black}}%
      \expandafter\def\csname LTa\endcsname{\color{black}}%
      \expandafter\def\csname LT0\endcsname{\color{black}}%
      \expandafter\def\csname LT1\endcsname{\color{black}}%
      \expandafter\def\csname LT2\endcsname{\color{black}}%
      \expandafter\def\csname LT3\endcsname{\color{black}}%
      \expandafter\def\csname LT4\endcsname{\color{black}}%
      \expandafter\def\csname LT5\endcsname{\color{black}}%
      \expandafter\def\csname LT6\endcsname{\color{black}}%
      \expandafter\def\csname LT7\endcsname{\color{black}}%
      \expandafter\def\csname LT8\endcsname{\color{black}}%
    \fi
  \fi
  \setlength{\unitlength}{0.0500bp}%
  \begin{picture}(4676.00,2834.00)%
    \gplgaddtomacro\gplbacktext{%
      \csname LTb\endcsname%
      \put(814,704){\makebox(0,0)[r]{\strut{}-12}}%
      \put(814,1077){\makebox(0,0)[r]{\strut{}-8}}%
      \put(814,1450){\makebox(0,0)[r]{\strut{}-4}}%
      \put(814,1823){\makebox(0,0)[r]{\strut{} 0}}%
      \put(814,2196){\makebox(0,0)[r]{\strut{} 4}}%
      \put(814,2569){\makebox(0,0)[r]{\strut{} 8}}%
      \put(1334,484){\makebox(0,0){\strut{} 0.2}}%
      \put(1978,484){\makebox(0,0){\strut{} 0.4}}%
      \put(2623,484){\makebox(0,0){\strut{} 0.6}}%
      \put(3267,484){\makebox(0,0){\strut{} 0.8}}%
      \put(3911,484){\makebox(0,0){\strut{} 1}}%
      \put(176,1636){\rotatebox{-270}{\makebox(0,0){\strut{}$\frac{E}{N}$}}}%
      \put(2612,154){\makebox(0,0){\strut{}$\frac{\phi}{2\pi}$}}%
    }%
    \gplgaddtomacro\gplfronttext{%
    }%
    \gplbacktext
    \put(0,0){\includegraphics{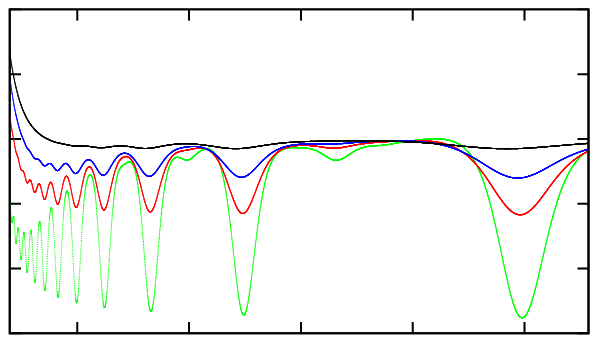}}%
    \gplfronttext
  \end{picture}%
\endgroup
\caption{The reduced energy per particle, $E/N$, in units of
  $d^2/(2\pi\epsilon_0R^3)$ as function of the angular distance,
  $\phi$, between the particles.  The different curves
  correspond to $h/R = 1.6, 1.0, 0.8$ and $0.6$ with smaller well
  depths for decreasing $h/R$.  The total particle number is $N= 100$.
 \label{fig:Eofa}}.
\end{figure}

\subsection{Equilibrium configurations}
Full coordinate variation of all positions imply minimization of an
$N$-dimensional function.  To simplify this problem we impose
constraints on the positions of the dipoles.  We first assume that all
the dipoles are equally spaced along the helix, that is the angular
distance, $ \phi = |\phi_i-\phi_j|$, between neighboring
particles is constant.  A given total number of particles, $N$, then
corresponds to varying the length of the helix.  The resulting energy per
dipole is shown in fig.~\ref{fig:Eofa} as a function of $ \phi$.
For sufficiently large $N$ the energy per particle is independent of the
length of the helix. 

The oscillating functions in fig.~\ref{fig:Eofa} are striking.  The
many minima at shorter and shorter angular distance correspond to an
integer number of dipoles per winding of the helix.  At larger
separation the oscillations continue but with cubic decrease of the
depths occurring at approximately integer multipla of $2\pi$.  The
depths of the minima are roughly equal for large $\phi$ while
decreasing towards smaller values of $\phi$.

To interpret these observations it is useful to recall that
head-to-tail vertical dipoles attract each other, while repulsion
occur between horizontal dipoles connected by a vector almost
perpendicular to the dipole moments.  The results in
fig.~\ref{fig:potential} then suggest the energy is dominated by the
attraction between the dipoles in different windings, since the
dominating minimum is close to $2\pi$.  The repulsion is much smaller
between dipoles on the same winding, where $\phi$ differ from $2\pi$.
Furthermore, the attractions between optimally placed dipoles on the
same winding decrease with decreasing values of $h/R$, see
fig.~\ref{fig:potential}.  These conclusions emerge because the energy per
dipole seems to be independent of the number of repulsions arising
from dipoles on the same winding.

With this overall interpretation we can look at more detailed
features.  The $h/R$ dependence of the positions of the minima is very
weak although visible.  The minimum at the highest dipole separation,
$\phi \approx 2\pi$, is in fact slightly less than $2\pi$.  This means
that each dipole in this case is not separated by exactly one winding,
and thus the positions of the $N$ dipoles for such a chain rotates
slowly around the axis of the helix. After about $75$ layers a full
revolution of $2\pi$ is completed.  As seen in fig.~\ref{fig:Eofa} the
deviation from $2\pi$ remains essentially unchanged by modest 
variation of $h/R$.  The second highest minimum position at around
$\pi$ in fig.~\ref{fig:Eofa} corresponds to two dipoles per
revolution.  The configuration is then dipoles forming two chains on
opposite sides of the helix as illustrated in
fig.~\ref{fig:draw2peromgang}.  Closer inspection reveals that each of
the two chains again slowly is rotating around the helix axis.

\begin{figure}
\begingroup
  \makeatletter
  \providecommand\color[2][]{%
    \GenericError{(gnuplot) \space\space\space\@spaces}{%
      Package color not loaded in conjunction with
      terminal option `colourtext'%
    }{See the gnuplot documentation for explanation.%
    }{Either use 'blacktext' in gnuplot or load the package
      color.sty in LaTeX.}%
    \renewcommand\color[2][]{}%
  }%
  \providecommand\includegraphics[2][]{%
    \GenericError{(gnuplot) \space\space\space\@spaces}{%
      Package graphicx or graphics not loaded%
    }{See the gnuplot documentation for explanation.%
    }{The gnuplot epslatex terminal needs graphicx.sty or graphics.sty.}%
    \renewcommand\includegraphics[2][]{}%
  }%
  \providecommand\rotatebox[2]{#2}%
  \@ifundefined{ifGPcolor}{%
    \newif\ifGPcolor
    \GPcolortrue
  }{}%
  \@ifundefined{ifGPblacktext}{%
    \newif\ifGPblacktext
    \GPblacktextfalse
  }{}%
  \let\gplgaddtomacro\g@addto@macro
  \gdef\gplbacktext{}%
  \gdef\gplfronttext{}%
  \makeatother
  \ifGPblacktext
    \def\colorrgb#1{}%
    \def\colorgray#1{}%
  \else
    \ifGPcolor
      \def\colorrgb#1{\color[rgb]{#1}}%
      \def\colorgray#1{\color[gray]{#1}}%
      \expandafter\def\csname LTw\endcsname{\color{white}}%
      \expandafter\def\csname LTb\endcsname{\color{black}}%
      \expandafter\def\csname LTa\endcsname{\color{black}}%
      \expandafter\def\csname LT0\endcsname{\color[rgb]{1,0,0}}%
      \expandafter\def\csname LT1\endcsname{\color[rgb]{0,1,0}}%
      \expandafter\def\csname LT2\endcsname{\color[rgb]{0,0,1}}%
      \expandafter\def\csname LT3\endcsname{\color[rgb]{1,0,1}}%
      \expandafter\def\csname LT4\endcsname{\color[rgb]{0,1,1}}%
      \expandafter\def\csname LT5\endcsname{\color[rgb]{1,1,0}}%
      \expandafter\def\csname LT6\endcsname{\color[rgb]{0,0,0}}%
      \expandafter\def\csname LT7\endcsname{\color[rgb]{1,0.3,0}}%
      \expandafter\def\csname LT8\endcsname{\color[rgb]{0.5,0.5,0.5}}%
    \else
      \def\colorrgb#1{\color{black}}%
      \def\colorgray#1{\color[gray]{#1}}%
      \expandafter\def\csname LTw\endcsname{\color{white}}%
      \expandafter\def\csname LTb\endcsname{\color{black}}%
      \expandafter\def\csname LTa\endcsname{\color{black}}%
      \expandafter\def\csname LT0\endcsname{\color{black}}%
      \expandafter\def\csname LT1\endcsname{\color{black}}%
      \expandafter\def\csname LT2\endcsname{\color{black}}%
      \expandafter\def\csname LT3\endcsname{\color{black}}%
      \expandafter\def\csname LT4\endcsname{\color{black}}%
      \expandafter\def\csname LT5\endcsname{\color{black}}%
      \expandafter\def\csname LT6\endcsname{\color{black}}%
      \expandafter\def\csname LT7\endcsname{\color{black}}%
      \expandafter\def\csname LT8\endcsname{\color{black}}%
    \fi
  \fi
  \setlength{\unitlength}{0.0500bp}%
  \begin{picture}(4676.00,2834.00)%
    \gplgaddtomacro\gplbacktext{%
      \csname LTb\endcsname%
      \put(946,704){\makebox(0,0)[r]{\strut{}-3}}%
      \put(946,1077){\makebox(0,0)[r]{\strut{}-2.8}}%
      \put(946,1450){\makebox(0,0)[r]{\strut{}-2.6}}%
      \put(946,1823){\makebox(0,0)[r]{\strut{}-2.4}}%
      \put(946,2196){\makebox(0,0)[r]{\strut{}-2.2}}%
      \put(946,2569){\makebox(0,0)[r]{\strut{}-2}}%
      \put(1078,484){\makebox(0,0){\strut{} 0.2}}%
      \put(1478,484){\makebox(0,0){\strut{} 0.4}}%
      \put(1878,484){\makebox(0,0){\strut{} 0.6}}%
      \put(2278,484){\makebox(0,0){\strut{} 0.8}}%
      \put(2679,484){\makebox(0,0){\strut{} 1}}%
      \put(3079,484){\makebox(0,0){\strut{} 1.2}}%
      \put(3479,484){\makebox(0,0){\strut{} 1.4}}%
      \put(3879,484){\makebox(0,0){\strut{} 1.6}}%
      \put(4279,484){\makebox(0,0){\strut{} 1.8}}%
      \put(176,1636){\rotatebox{-270}{\makebox(0,0){\strut{}$\left(\frac{h}{R}\right)^3\frac{E}{N}$}}}%
      \put(2678,154){\makebox(0,0){\strut{}$\frac{h}{R}$}}%
    }%
    \gplgaddtomacro\gplfronttext{%
    }%
    \gplbacktext
    \put(0,0){\includegraphics{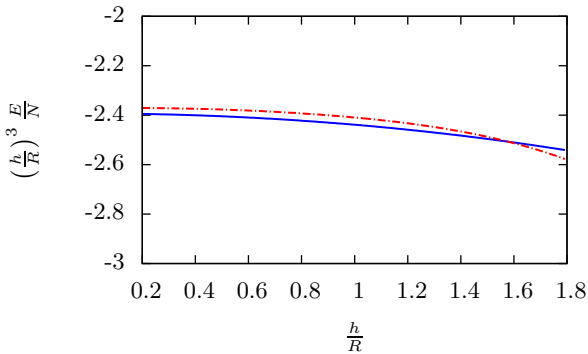}}%
    \gplfronttext
  \end{picture}%
\endgroup
\caption{Reduced energy per dipole multiplied by $(h/R)^3$ as a
  function of the ratio $\frac{h}{R}$ for one (full blue) and three
  (dashed red) dipoles per revolution.  \label{fig:EofhR}}
\end{figure}

The energy shown in fig.~\ref{fig:Eofa} depends on the azimuthal angle
and the ratio $h/R$ where the latter in principle can vary from $0$ to
$\pi \sqrt{2} \approx 4.4$, For two dipoles separated by $r$ in three
dimensions the energy scales as $r^{-3}$.  For two dipoles separated
by one winding the three dimensional distance scales as $h$, and
according to Eq.(\ref{eq:potential}) the energy per dipole must scale
as $h^{-3}$.  If this in-chain interaction is the dominating
contribution, the energy per dipole should scale in this way.  This is
another way to investigate the above conclusion that the attraction
between dipoles in different windings dominate the total energy, even
for higher dipole density.

The expected scaling is tested in fig.~\ref{fig:EofhR}, where we show
the product $\frac{E}{N}\left(\frac{h}{R}\right)^3$, which should be
constant under these assumptions of in-chain dominance.  Over the
range of $h/R$ from $0.2$ to $1.8$ the deviations from an average
constant is less than $10\%$.  This change is mostly due to the slight
rotation of equilibrium angle with each revolution, i.e. dependency on
both $h/R$ and the angle.  The in-chain dominance would be violated by
even larger values of $h$ where out-of-chain and in-chain distances
become comparable. However, configurations with too large $h$-values
are unstable and hence uninteresting.

The opposite limit of $h\rightarrow 0$ would also violate the
distinction between particles in different chains as evident from the
geometry where all particles would be on a circle and all repelling
each other by an $h$-independent interaction.  The equilibrium
structure then necessarily must be equidistant spacing around the
resulting ring for all particle numbers.  The chain structure has
disappeared.  However, this structure is only reached in the extreme
limit of $h=0$, while the chain structure is maintained for any
finite, even infinitesimally small, value of $h/R$.  The present
calculations assume equidistant angular spacing, which implies that
the in-chain particles approach each other in head-to-tail attractive
configurations, and simultaneously all out-of-chain interactions
become repulsive.  Then the energy scaling of these rather unstable
configurations approach $h^{-3}$.

The understanding of equilibrium configurations as effective chain
structures has some interesting consequences for the spectrum of
normal modes.

\section{Normal modes} 
Vibrations around equilibrium configurations reveal intrinsic
properties of the vibrating structures.  The corresponding uncoupled
motion of orthogonal modes can be found by expansion of the energy in
Eq.(\ref{eq:totalenergy}) to second order in coordinate deviations
from the set of their equilibrium values $\{\phi_j^{(0)}\}$. Then we have
\begin{eqnarray}  \label{eq:deriv2}
 E-E_0 &\approx& \frac{1}{2} \sum_{i,j} K_{i,j}
 (\phi_i-\phi_i^{(0)})  (\phi_j-\phi_j^{(0)})\;,  \\  \label{eq:deriv2K} 
 K_{i,j} &=& \frac{\partial^2 E}{\partial \phi_i \partial \phi_j} \;,
\end{eqnarray}
where $E_0$ is the energy of the equilibrium configuration.  Both the
vanishing (omitted) first and the second derivatives ($K$) are taken
at the equilibrium.  The variable change corresponding to
diagonalizing the matrix $K$ in Eq.(\ref{eq:deriv2K}) provides a set
of eigenvalues defined by $ \frac{1}{2} m \omega_n^2$.  The resulting
frequencies $\omega^2_n$ are all positive if the equilibrium
configuration is stable.  Conversely, if one or more eigenvalues,
$\omega^2_n$, are negative, then the system is unstable towards
coordinate deviations in the corresponding directions.

\subsection{Stability} 
We calculate the elements of the $K$-matrix in Eq.(\ref{eq:deriv2})
analytically from the expression in Eq.(\ref{eq:potential}).  We
subsequently diagonalize this matrix and obtain the eigenvalues
expressed as eigenfrequencies.  The corresponding eigenvectors include
all the $N$ angular coordinates, in contrast to the energy
calculations in the previous section where constant differences
between neighbouring particles were kept.  This extension of configuration
space implies that the computed eigenfrequencies can show instability
corresponding to vibrational modes different from vibrations where the
nearest neighbour distances are kept.

We emphasize that a near (first order) miss of the minimum due to a
too restricted (constant difference) variational space do not
jeopardize the stability condition since the second derivatives remain
unchanged by moving from approximate to true (close-lying) minimum.  A
conclusion of stability (all $\omega^2_n$ are positive) of the
calculated restricted solution then means that the marginally
deviating true solution also is stable.  On the other hand,
instability of the restricted solution implies that the true solution
may deviate substantially.

It turns out that the restricted configurations (constant angular
separation) are not always stable, and that the stability depends, on
both the geometry of the helix ($\frac{h}{R}$), and the total number of
dipoles $N$.  The configuration with one dipole per winding is always
stable as long as the short range interaction is repulsive, that is
$\frac{h}{R}<\pi \sqrt{2}$.  For two, three and more dipoles per
winding the picture gets more complicated.

It is tempting to believe that distances corresponding to the minima
in fig.~\ref{fig:Eofa} are good candidates for stable configurations.
We therefore place the dipoles at these points with equidistant
separation such that two, three and four dipoles are at $\phi \approx
\pi$, $ \phi \approx 2 \pi/3$, and $\phi \approx \pi/2$, respectively.
We then increase the total number of particles, $N$, or equivalently
the extension of the helix, and observe when, or possibly if, these
configurations become unstable.  Obviously, an unstable configuration
must require another structure of lower energy.  However, finding
those structures for increasing $N$ quickly becomes a complicated
many-body problem.  These solutions are in general beyond the scope of
this paper.

\begin{figure}
\begingroup
  \makeatletter
  \providecommand\color[2][]{%
    \GenericError{(gnuplot) \space\space\space\@spaces}{%
      Package color not loaded in conjunction with
      terminal option `colourtext'%
    }{See the gnuplot documentation for explanation.%
    }{Either use 'blacktext' in gnuplot or load the package
      color.sty in LaTeX.}%
    \renewcommand\color[2][]{}%
  }%
  \providecommand\includegraphics[2][]{%
    \GenericError{(gnuplot) \space\space\space\@spaces}{%
      Package graphicx or graphics not loaded%
    }{See the gnuplot documentation for explanation.%
    }{The gnuplot epslatex terminal needs graphicx.sty or graphics.sty.}%
    \renewcommand\includegraphics[2][]{}%
  }%
  \providecommand\rotatebox[2]{#2}%
  \@ifundefined{ifGPcolor}{%
    \newif\ifGPcolor
    \GPcolortrue
  }{}%
  \@ifundefined{ifGPblacktext}{%
    \newif\ifGPblacktext
    \GPblacktextfalse
  }{}%
  \let\gplgaddtomacro\g@addto@macro
  \gdef\gplbacktext{}%
  \gdef\gplfronttext{}%
  \makeatother
  \ifGPblacktext
    \def\colorrgb#1{}%
    \def\colorgray#1{}%
  \else
    \ifGPcolor
      \def\colorrgb#1{\color[rgb]{#1}}%
      \def\colorgray#1{\color[gray]{#1}}%
      \expandafter\def\csname LTw\endcsname{\color{white}}%
      \expandafter\def\csname LTb\endcsname{\color{black}}%
      \expandafter\def\csname LTa\endcsname{\color{black}}%
      \expandafter\def\csname LT0\endcsname{\color[rgb]{1,0,0}}%
      \expandafter\def\csname LT1\endcsname{\color[rgb]{0,1,0}}%
      \expandafter\def\csname LT2\endcsname{\color[rgb]{0,0,1}}%
      \expandafter\def\csname LT3\endcsname{\color[rgb]{1,0,1}}%
      \expandafter\def\csname LT4\endcsname{\color[rgb]{0,1,1}}%
      \expandafter\def\csname LT5\endcsname{\color[rgb]{1,1,0}}%
      \expandafter\def\csname LT6\endcsname{\color[rgb]{0,0,0}}%
      \expandafter\def\csname LT7\endcsname{\color[rgb]{1,0.3,0}}%
      \expandafter\def\csname LT8\endcsname{\color[rgb]{0.5,0.5,0.5}}%
    \else
      \def\colorrgb#1{\color{black}}%
      \def\colorgray#1{\color[gray]{#1}}%
      \expandafter\def\csname LTw\endcsname{\color{white}}%
      \expandafter\def\csname LTb\endcsname{\color{black}}%
      \expandafter\def\csname LTa\endcsname{\color{black}}%
      \expandafter\def\csname LT0\endcsname{\color{black}}%
      \expandafter\def\csname LT1\endcsname{\color{black}}%
      \expandafter\def\csname LT2\endcsname{\color{black}}%
      \expandafter\def\csname LT3\endcsname{\color{black}}%
      \expandafter\def\csname LT4\endcsname{\color{black}}%
      \expandafter\def\csname LT5\endcsname{\color{black}}%
      \expandafter\def\csname LT6\endcsname{\color{black}}%
      \expandafter\def\csname LT7\endcsname{\color{black}}%
      \expandafter\def\csname LT8\endcsname{\color{black}}%
    \fi
  \fi
  \setlength{\unitlength}{0.0500bp}%
  \begin{picture}(4676.00,2834.00)%
    \gplgaddtomacro\gplbacktext{%
      \csname LTb\endcsname%
      \put(1078,704){\makebox(0,0)[r]{\strut{} 0}}%
      \put(1078,1077){\makebox(0,0)[r]{\strut{} 400}}%
      \put(1078,1450){\makebox(0,0)[r]{\strut{} 800}}%
      \put(1078,1823){\makebox(0,0)[r]{\strut{} 1200}}%
      \put(1078,2196){\makebox(0,0)[r]{\strut{} 1600}}%
      \put(1078,2569){\makebox(0,0)[r]{\strut{} 2000}}%
      \put(1210,484){\makebox(0,0){\strut{} 0.4}}%
      \put(1722,484){\makebox(0,0){\strut{} 0.6}}%
      \put(2233,484){\makebox(0,0){\strut{} 0.8}}%
      \put(2745,484){\makebox(0,0){\strut{} 1}}%
      \put(3256,484){\makebox(0,0){\strut{} 1.2}}%
      \put(3767,484){\makebox(0,0){\strut{} 1.4}}%
      \put(4279,484){\makebox(0,0){\strut{} 1.6}}%
      \put(176,1636){\rotatebox{-270}{\makebox(0,0){\strut{}$N_{crit}$}}}%
      \put(2744,154){\makebox(0,0){\strut{}$\frac{h}{R}$}}%
    }%
    \gplgaddtomacro\gplfronttext{%
    }%
    \gplbacktext
    \put(0,0){\includegraphics{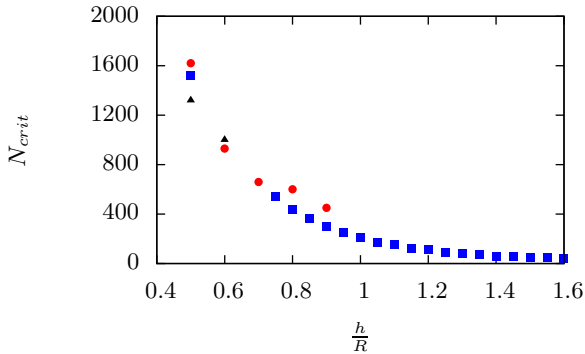}}%
    \gplfronttext
  \end{picture}%
\endgroup
\caption{The critical number of dipoles $N_{crit}$ as a function of
  $\frac{h}{R}$.  For $N>N _{crit}$ the equidistant configuration
  becomes unstable.  The blue square is for two dipoles per winding,
  the red circle is with three dipoles per winding and the black triangle is for 4 dipoles per winding\label{fig:Ncrit}}
\end{figure}

In fig.~\ref{fig:Ncrit}, the critical number of dipoles $N_{crit}$, is
shown as a function of $\frac{h}{R}$.  Values of $N$ smaller (larger)
than $N_{crit}$ are then stable (unstable).  If there is no point for
a particular value of $\frac{h}{R}$, it is because the system is
stable up to at least our maximum total number, $N=1600$.  Two dipoles
per winding are stable in the equidistant configuration when
$\frac{h}{R}<0.65$, at least up to $N = 1600$. For $\frac{h}{R}>0.65$
the system becomes unstable at much smaller values of $N$ decreasing
from about $550$ towards zero as $\frac{h}{R}$ increases.  In
contrast, for three dipoles per winding the critical value of $N$
decreases from about $1600$ to about $500$ when $\frac{h}{R}$
increases from $0.5$ to $0.9$. For $\frac{h}{R} > 0.9$, the
equidistant configuration is stable for three dipoles per winding for
all $N < 1200$.

The stability behaviour for three and four dipoles per winding is
similar but stability arises already for larger values than $h/R
\approx 0.6$.  This smaller threshold of $h$-instability has to arise
from the presence of the fourth particle.  Roughly in the interval,
$0.2 < h/R < 0.6$, the four equidistantly placed dipoles become
unstable only for relatively large dipole number as shown in
fig.~\ref{fig:Ncrit}. For small $h/R$-values, the behavior is similar
to our three cases in fig.~\ref{fig:Ncrit}.  The apparent extra
stability of four chains are due to a strong repulsion between
neighboring dipoles in the same circle.  If possible, they lock each
other in the minima seen in fig.~\ref{fig:Eofa}.

When $h/R $ decreases below about $0.2$ the equidistant four-particle
system again becomes stable, but now due to the overwhelmingly
dominating in-chain head-to-tail attraction.  This configuration is
reached by decreasing $h$ towards zero for equidistant angular
distances.  The stability in at least a local minimum is guaranteed
for any particle number per winding by strong in-chain attraction and
a comparably much smaller out-of-chain repulsion.  The stability
behavior for different dipole numbers is a sobering illustration of
the possible many-body complexity of even the present one-dimensional
system.  The detailed structure-changes responsible for these
instabilities are investigated in the next subsection.

\begin{figure}
\begingroup
  \makeatletter
  \providecommand\color[2][]{%
    \GenericError{(gnuplot) \space\space\space\@spaces}{%
      Package color not loaded in conjunction with
      terminal option `colourtext'%
    }{See the gnuplot documentation for explanation.%
    }{Either use 'blacktext' in gnuplot or load the package
      color.sty in LaTeX.}%
    \renewcommand\color[2][]{}%
  }%
  \providecommand\includegraphics[2][]{%
    \GenericError{(gnuplot) \space\space\space\@spaces}{%
      Package graphicx or graphics not loaded%
    }{See the gnuplot documentation for explanation.%
    }{The gnuplot epslatex terminal needs graphicx.sty or graphics.sty.}%
    \renewcommand\includegraphics[2][]{}%
  }%
  \providecommand\rotatebox[2]{#2}%
  \@ifundefined{ifGPcolor}{%
    \newif\ifGPcolor
    \GPcolortrue
  }{}%
  \@ifundefined{ifGPblacktext}{%
    \newif\ifGPblacktext
    \GPblacktextfalse
  }{}%
  \let\gplgaddtomacro\g@addto@macro
  \gdef\gplbacktext{}%
  \gdef\gplfronttext{}%
  \makeatother
  \ifGPblacktext
    \def\colorrgb#1{}%
    \def\colorgray#1{}%
  \else
    \ifGPcolor
      \def\colorrgb#1{\color[rgb]{#1}}%
      \def\colorgray#1{\color[gray]{#1}}%
      \expandafter\def\csname LTw\endcsname{\color{white}}%
      \expandafter\def\csname LTb\endcsname{\color{black}}%
      \expandafter\def\csname LTa\endcsname{\color{black}}%
      \expandafter\def\csname LT0\endcsname{\color[rgb]{1,0,0}}%
      \expandafter\def\csname LT1\endcsname{\color[rgb]{0,1,0}}%
      \expandafter\def\csname LT2\endcsname{\color[rgb]{0,0,1}}%
      \expandafter\def\csname LT3\endcsname{\color[rgb]{1,0,1}}%
      \expandafter\def\csname LT4\endcsname{\color[rgb]{0,1,1}}%
      \expandafter\def\csname LT5\endcsname{\color[rgb]{1,1,0}}%
      \expandafter\def\csname LT6\endcsname{\color[rgb]{0,0,0}}%
      \expandafter\def\csname LT7\endcsname{\color[rgb]{1,0.3,0}}%
      \expandafter\def\csname LT8\endcsname{\color[rgb]{0.5,0.5,0.5}}%
    \else
      \def\colorrgb#1{\color{black}}%
      \def\colorgray#1{\color[gray]{#1}}%
      \expandafter\def\csname LTw\endcsname{\color{white}}%
      \expandafter\def\csname LTb\endcsname{\color{black}}%
      \expandafter\def\csname LTa\endcsname{\color{black}}%
      \expandafter\def\csname LT0\endcsname{\color{black}}%
      \expandafter\def\csname LT1\endcsname{\color{black}}%
      \expandafter\def\csname LT2\endcsname{\color{black}}%
      \expandafter\def\csname LT3\endcsname{\color{black}}%
      \expandafter\def\csname LT4\endcsname{\color{black}}%
      \expandafter\def\csname LT5\endcsname{\color{black}}%
      \expandafter\def\csname LT6\endcsname{\color{black}}%
      \expandafter\def\csname LT7\endcsname{\color{black}}%
      \expandafter\def\csname LT8\endcsname{\color{black}}%
    \fi
  \fi
  \setlength{\unitlength}{0.0500bp}%
  \begin{picture}(4676.00,2834.00)%
    \gplgaddtomacro\gplbacktext{%
      \csname LTb\endcsname%
      \put(1210,704){\makebox(0,0)[r]{\strut{}-4.890}}%
      \put(1210,1077){\makebox(0,0)[r]{\strut{}-4.888}}%
      \put(1210,1450){\makebox(0,0)[r]{\strut{}-4.886}}%
      \put(1210,1823){\makebox(0,0)[r]{\strut{}-4.884}}%
      \put(1210,2196){\makebox(0,0)[r]{\strut{}-4.882}}%
      \put(1210,2569){\makebox(0,0)[r]{\strut{}-4.880}}%
      \put(1342,484){\makebox(0,0){\strut{} 0}}%
      \put(1957,484){\makebox(0,0){\strut{} 0.2}}%
      \put(2572,484){\makebox(0,0){\strut{} 0.4}}%
      \put(3187,484){\makebox(0,0){\strut{} 0.6}}%
      \put(3802,484){\makebox(0,0){\strut{} 0.8}}%
      \put(176,1636){\rotatebox{-270}{\makebox(0,0){\strut{}$\frac{E}{N}$}}}%
      \put(2810,154){\makebox(0,0){\strut{}$\frac{\phi}{2\pi}$}}%
    }%
    \gplgaddtomacro\gplfronttext{%
    }%
    \gplbacktext
    \put(0,0){\includegraphics{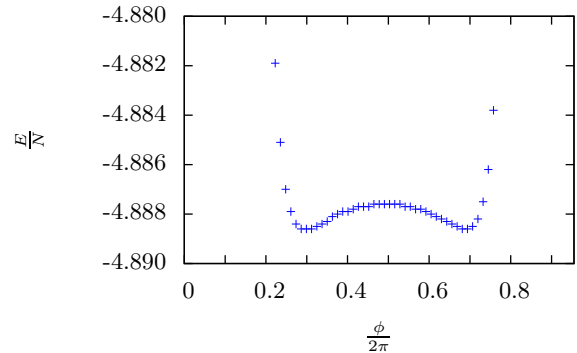}}%
    \gplfronttext
  \end{picture}%
\endgroup
\caption{The reduced energy per dipole, $E/N$, as function of the
  angular distance between two chains of dipoles for $N=500$ and $h =R$.  
  \label{fig:EofChain}}
\end{figure}

\subsection{Chain structure}

To understand the computed properties we first return to the two chain
structure illustrated in fig.~\ref{fig:draw2peromgang} for two dipoles
per winding.  We focus on two nearest neighbour dipoles belonging to
different chains.  When $\frac{h}{R}$ increases for given $N$ their
interaction changes from repulsive to attractive, and therefore it
becomes energetically unfavourable to maintain the configuration of
equal angular spacing.  It is advantageoues to move the two chains
closer to each other. For example, for $N \approx 600$ the stable
configuration for small $\frac{h}{R}$ becomes unstable when
$\frac{h}{R}$ increases above $0.75$.

This result can be understood by extending the helix by adding one
pair at a time.  Each of these extra dipoles at the end of the helix
repel the neighbor in the opposite chain in the circle beneath.  In
contrast, the interaction to dipoles on circles far away would be
attractive because such pairs approach the head-to-tail attractive
configurations.  The larger $N$ the more of these attractions which
eventually would dominate over the repulsion between nearest
neighbors.  This is directly related to the two-body potential shown
in fig.~\ref{fig:potential} where the attraction appears already for
$\phi > 2\pi$.

This explanation is tested by varying the distance between the two
chains in fig.~\ref{fig:draw2peromgang}.  The resulting energy per
dipole is shown in fig.~\ref{fig:EofChain}, where the equidistant
configuration corresponds to the angle close to $\pi$.  The
corresponding maximum demonstrate instability and the two minima at
about $1.8$ and $4.3$ strongly suggest an equilibrium configuration
where the two chains remain but at a smaller distance.  The two minima
are equivalent as each chain can be shifted by moving points either up
or down the helix.  A further confirmation of this two-chain
interpretation for equidistant chains is derived from the fact that
the first unstable normal mode with the imaginary frequency
corresponds to the structure of two chains vibrating against each
other.  All other normal modes of equidistant chain vibrations reveal
stability.  The calculated normal mode frequencies for the new
non-equidistant minimum configurations all show stability for large $N$.

\begin{figure}
\begingroup
  \makeatletter
  \providecommand\color[2][]{%
    \GenericError{(gnuplot) \space\space\space\@spaces}{%
      Package color not loaded in conjunction with
      terminal option `colourtext'%
    }{See the gnuplot documentation for explanation.%
    }{Either use 'blacktext' in gnuplot or load the package
      color.sty in LaTeX.}%
    \renewcommand\color[2][]{}%
  }%
  \providecommand\includegraphics[2][]{%
    \GenericError{(gnuplot) \space\space\space\@spaces}{%
      Package graphicx or graphics not loaded%
    }{See the gnuplot documentation for explanation.%
    }{The gnuplot epslatex terminal needs graphicx.sty or graphics.sty.}%
    \renewcommand\includegraphics[2][]{}%
  }%
  \providecommand\rotatebox[2]{#2}%
  \@ifundefined{ifGPcolor}{%
    \newif\ifGPcolor
    \GPcolortrue
  }{}%
  \@ifundefined{ifGPblacktext}{%
    \newif\ifGPblacktext
    \GPblacktextfalse
  }{}%
  \let\gplgaddtomacro\g@addto@macro
  \gdef\gplbacktext{}%
  \gdef\gplfronttext{}%
  \makeatother
  \ifGPblacktext
    \def\colorrgb#1{}%
    \def\colorgray#1{}%
  \else
    \ifGPcolor
      \def\colorrgb#1{\color[rgb]{#1}}%
      \def\colorgray#1{\color[gray]{#1}}%
      \expandafter\def\csname LTw\endcsname{\color{white}}%
      \expandafter\def\csname LTb\endcsname{\color{black}}%
      \expandafter\def\csname LTa\endcsname{\color{black}}%
      \expandafter\def\csname LT0\endcsname{\color[rgb]{1,0,0}}%
      \expandafter\def\csname LT1\endcsname{\color[rgb]{0,1,0}}%
      \expandafter\def\csname LT2\endcsname{\color[rgb]{0,0,1}}%
      \expandafter\def\csname LT3\endcsname{\color[rgb]{1,0,1}}%
      \expandafter\def\csname LT4\endcsname{\color[rgb]{0,1,1}}%
      \expandafter\def\csname LT5\endcsname{\color[rgb]{1,1,0}}%
      \expandafter\def\csname LT6\endcsname{\color[rgb]{0,0,0}}%
      \expandafter\def\csname LT7\endcsname{\color[rgb]{1,0.3,0}}%
      \expandafter\def\csname LT8\endcsname{\color[rgb]{0.5,0.5,0.5}}%
    \else
      \def\colorrgb#1{\color{black}}%
      \def\colorgray#1{\color[gray]{#1}}%
      \expandafter\def\csname LTw\endcsname{\color{white}}%
      \expandafter\def\csname LTb\endcsname{\color{black}}%
      \expandafter\def\csname LTa\endcsname{\color{black}}%
      \expandafter\def\csname LT0\endcsname{\color{black}}%
      \expandafter\def\csname LT1\endcsname{\color{black}}%
      \expandafter\def\csname LT2\endcsname{\color{black}}%
      \expandafter\def\csname LT3\endcsname{\color{black}}%
      \expandafter\def\csname LT4\endcsname{\color{black}}%
      \expandafter\def\csname LT5\endcsname{\color{black}}%
      \expandafter\def\csname LT6\endcsname{\color{black}}%
      \expandafter\def\csname LT7\endcsname{\color{black}}%
      \expandafter\def\csname LT8\endcsname{\color{black}}%
    \fi
  \fi
  \setlength{\unitlength}{0.0500bp}%
  \begin{picture}(4676.00,4676.00)%
    \gplgaddtomacro\gplbacktext{%
      \csname LTb\endcsname%
      \put(1210,704){\makebox(0,0)[r]{\strut{}-33.580}}%
      \put(1210,1167){\makebox(0,0)[r]{\strut{}-33.57}}%
      \put(1210,1631){\makebox(0,0)[r]{\strut{}-33.56}}%
      \put(1210,2094){\makebox(0,0)[r]{\strut{}-33.55}}%
      \put(1210,2557){\makebox(0,0)[r]{\strut{}-33.54}}%
      \put(1210,3021){\makebox(0,0)[r]{\strut{}-33.53}}%
      \put(1210,3484){\makebox(0,0)[r]{\strut{}-33.52}}%
      \put(1210,3948){\makebox(0,0)[r]{\strut{}-33.51}}%
      \put(1210,4411){\makebox(0,0)[r]{\strut{}-33.50}}%
      \put(1468,484){\makebox(0,0){\strut{} 0.1}}%
      \put(2083,484){\makebox(0,0){\strut{} 0.2}}%
      \put(2698,484){\makebox(0,0){\strut{} 0.3}}%
      \put(3313,484){\makebox(0,0){\strut{} 0.4}}%
      \put(3928,484){\makebox(0,0){\strut{} 0.5}}%
      \put(176,2557){\rotatebox{-270}{\makebox(0,0){\strut{}$\frac{E}{N}$}}}%
      \put(2810,154){\makebox(0,0){\strut{}$\frac{\phi}{2*\pi}$}}%
    }%
    \gplgaddtomacro\gplfronttext{%
    }%
    \gplbacktext
    \put(0,0){\includegraphics{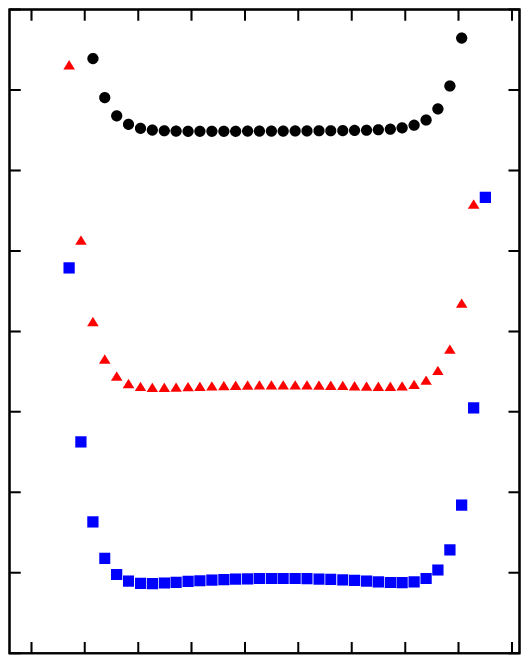}}%
    \gplfronttext
  \end{picture}%
\endgroup
\caption{Reduced energy per dipole, $\frac{E}{N}$, as a function of
  the position of a third chain of dipoles between two other chains,
  the first two chains are separated by $\phi=\frac{2}{3}2\pi$
  radians, and the position of the third is measured from the
  first. This is done from top to bottom for $N=500$ (black circle)
  $N=700$ (red triangle) and $N=1000$ (blue square) dipoles and
  $\frac{h}{R}=0.6$. \label{fig:h06}}
\end{figure} 

We continue to test the idea of chains as the basic ingredients and
explanations of the stability results in fig.~\ref{fig:Ncrit} where
three chains equidistantly placed on the helix becomes unstable for
$N> N_{crit} \approx 900$ and $\frac{h}{R}=0.6$.  The properties of this
instability can be investigated by a procedure similar to that leading
to fig.~\ref{fig:EofChain}.  We first form two chains with an angular
distance of $\phi = \frac{2}{3}2\pi$ corresponding to the third
minimum in fig.~\ref{fig:Eofa}. If a third chain of dipoles is placed
directly in between the first two we recover the equidistant
configuration for three dipoles per winding.  Instead we vary the
position of the third chain between the first two chains.  

The resulting energies are shown in fig.~\ref{fig:h06} for three
different total number of particles.  For $N=500$ we see the minimum
at $\phi = \frac{2}{3}2\pi$, that is directly between the two first
chains corresponding to the equidistant configuration, and consistent
with $N < N_{crit} \approx 900$.  For $N=700$ it is difficult to
extract an accurate value for the position of the minimum, but
increasing further to $N=1000$ two separate minima appear on opposite
site of the value for the (now unstable) equidistant configuration.

The interpretation is again supported by the observation that the
structure of the first unstable normal mode with imaginary frequency
corresponds to the two nearest chains vibrating against each other,
while all other normal modes are stable.  All these properties show
that even when the equidistant configuration becomes unstable, the
chain structure can still be considered the basic constituent.  This
means that the attraction of two dipoles above each other in a chain
dominates over the interaction between dipoles in different chains.

\begin{figure*}
\centering
\begingroup
  \makeatletter
  \providecommand\color[2][]{%
    \GenericError{(gnuplot) \space\space\space\@spaces}{%
      Package color not loaded in conjunction with
      terminal option `colourtext'%
    }{See the gnuplot documentation for explanation.%
    }{Either use 'blacktext' in gnuplot or load the package
      color.sty in LaTeX.}%
    \renewcommand\color[2][]{}%
  }%
  \providecommand\includegraphics[2][]{%
    \GenericError{(gnuplot) \space\space\space\@spaces}{%
      Package graphicx or graphics not loaded%
    }{See the gnuplot documentation for explanation.%
    }{The gnuplot epslatex terminal needs graphicx.sty or graphics.sty.}%
    \renewcommand\includegraphics[2][]{}%
  }%
  \providecommand\rotatebox[2]{#2}%
  \@ifundefined{ifGPcolor}{%
    \newif\ifGPcolor
    \GPcolortrue
  }{}%
  \@ifundefined{ifGPblacktext}{%
    \newif\ifGPblacktext
    \GPblacktextfalse
  }{}%
  \let\gplgaddtomacro\g@addto@macro
  \gdef\gplbacktext{}%
  \gdef\gplfronttext{}%
  \makeatother
  \ifGPblacktext
    \def\colorrgb#1{}%
    \def\colorgray#1{}%
  \else
    \ifGPcolor
      \def\colorrgb#1{\color[rgb]{#1}}%
      \def\colorgray#1{\color[gray]{#1}}%
      \expandafter\def\csname LTw\endcsname{\color{white}}%
      \expandafter\def\csname LTb\endcsname{\color{black}}%
      \expandafter\def\csname LTa\endcsname{\color{black}}%
      \expandafter\def\csname LT0\endcsname{\color[rgb]{1,0,0}}%
      \expandafter\def\csname LT1\endcsname{\color[rgb]{0,1,0}}%
      \expandafter\def\csname LT2\endcsname{\color[rgb]{0,0,1}}%
      \expandafter\def\csname LT3\endcsname{\color[rgb]{1,0,1}}%
      \expandafter\def\csname LT4\endcsname{\color[rgb]{0,1,1}}%
      \expandafter\def\csname LT5\endcsname{\color[rgb]{1,1,0}}%
      \expandafter\def\csname LT6\endcsname{\color[rgb]{0,0,0}}%
      \expandafter\def\csname LT7\endcsname{\color[rgb]{1,0.3,0}}%
      \expandafter\def\csname LT8\endcsname{\color[rgb]{0.5,0.5,0.5}}%
    \else
      \def\colorrgb#1{\color{black}}%
      \def\colorgray#1{\color[gray]{#1}}%
      \expandafter\def\csname LTw\endcsname{\color{white}}%
      \expandafter\def\csname LTb\endcsname{\color{black}}%
      \expandafter\def\csname LTa\endcsname{\color{black}}%
      \expandafter\def\csname LT0\endcsname{\color{black}}%
      \expandafter\def\csname LT1\endcsname{\color{black}}%
      \expandafter\def\csname LT2\endcsname{\color{black}}%
      \expandafter\def\csname LT3\endcsname{\color{black}}%
      \expandafter\def\csname LT4\endcsname{\color{black}}%
      \expandafter\def\csname LT5\endcsname{\color{black}}%
      \expandafter\def\csname LT6\endcsname{\color{black}}%
      \expandafter\def\csname LT7\endcsname{\color{black}}%
      \expandafter\def\csname LT8\endcsname{\color{black}}%
    \fi
  \fi
  \setlength{\unitlength}{0.0500bp}%
  \begin{picture}(6236.00,5952.00)%
    \gplgaddtomacro\gplbacktext{%
      \csname LTb\endcsname%
      \put(491,3023){\makebox(0,0)[r]{\strut{} 0}}%
      \put(491,3489){\makebox(0,0)[r]{\strut{} 10}}%
      \put(491,3956){\makebox(0,0)[r]{\strut{} 20}}%
      \put(491,4422){\makebox(0,0)[r]{\strut{} 30}}%
      \put(491,4889){\makebox(0,0)[r]{\strut{} 40}}%
      \put(491,5355){\makebox(0,0)[r]{\strut{} 50}}%
      \put(623,2756){\makebox(0,0){\strut{} 0}}%
      \put(1390,2756){\makebox(0,0){\strut{} 2}}%
      \put(2158,2756){\makebox(0,0){\strut{} 4}}%
      \put(2925,2756){\makebox(0,0){\strut{} 6}}%
      \put(3693,2756){\makebox(0,0){\strut{} 8}}%
      \put(4460,2756){\makebox(0,0){\strut{} 10}}%
      \put(5227,2756){\makebox(0,0){\strut{} 12}}%
      \put(-147,4165){\rotatebox{-270}{\makebox(0,0){\strut{}$\omega^2$}}}%
      \put(3117,2470){\makebox(0,0){\strut{}}}%
      \put(3117,5685){\makebox(0,0){\strut{}Eigen mode}}%
    }%
    \gplgaddtomacro\gplfronttext{%
    }%
    \gplgaddtomacro\gplbacktext{%
      \csname LTb\endcsname%
      \put(491,595){\makebox(0,0)[r]{\strut{}-1}}%
      \put(491,967){\makebox(0,0)[r]{\strut{}-0.5}}%
      \put(491,1339){\makebox(0,0)[r]{\strut{} 0}}%
      \put(491,1710){\makebox(0,0)[r]{\strut{} 0.5}}%
      \put(491,2082){\makebox(0,0)[r]{\strut{} 1}}%
      \put(623,375){\makebox(0,0){\strut{} 0}}%
      \put(831,375){\makebox(0,0){\strut{} 2}}%
      \put(1039,375){\makebox(0,0){\strut{} 4}}%
      \put(1247,375){\makebox(0,0){\strut{} 6}}%
      \put(1454,375){\makebox(0,0){\strut{} 8}}%
      \put(1662,375){\makebox(0,0){\strut{} 10}}%
      \put(1870,375){\makebox(0,0){\strut{} 12}}%
      \put(-279,1338){\rotatebox{-270}{\makebox(0,0){\strut{}Amplitude}}}%
      \put(1246,89){\makebox(0,0){\strut{}}}%
    }%
    \gplgaddtomacro\gplfronttext{%
    }%
    \gplgaddtomacro\gplbacktext{%
      \csname LTb\endcsname%
      \put(2362,595){\makebox(0,0)[r]{\strut{}-1}}%
      \put(2362,967){\makebox(0,0)[r]{\strut{}-0.5}}%
      \put(2362,1339){\makebox(0,0)[r]{\strut{} 0}}%
      \put(2362,1710){\makebox(0,0)[r]{\strut{} 0.5}}%
      \put(2362,2082){\makebox(0,0)[r]{\strut{} 1}}%
      \put(2494,375){\makebox(0,0){\strut{} 0}}%
      \put(2702,375){\makebox(0,0){\strut{} 2}}%
      \put(2910,375){\makebox(0,0){\strut{} 4}}%
      \put(3118,375){\makebox(0,0){\strut{} 6}}%
      \put(3325,375){\makebox(0,0){\strut{} 8}}%
      \put(3533,375){\makebox(0,0){\strut{} 10}}%
      \put(3741,375){\makebox(0,0){\strut{} 12}}%
      \put(1812,1338){\rotatebox{-270}{\makebox(0,0){\strut{}}}}%
      \put(3117,45){\makebox(0,0){\strut{}Dipole number along the helix}}%
    }%
    \gplgaddtomacro\gplfronttext{%
    }%
    \gplgaddtomacro\gplbacktext{%
      \csname LTb\endcsname%
      \put(4233,595){\makebox(0,0)[r]{\strut{}-1}}%
      \put(4233,967){\makebox(0,0)[r]{\strut{}-0.5}}%
      \put(4233,1339){\makebox(0,0)[r]{\strut{} 0}}%
      \put(4233,1710){\makebox(0,0)[r]{\strut{} 0.5}}%
      \put(4233,2082){\makebox(0,0)[r]{\strut{} 1}}%
      \put(4365,375){\makebox(0,0){\strut{} 0}}%
      \put(4573,375){\makebox(0,0){\strut{} 2}}%
      \put(4780,375){\makebox(0,0){\strut{} 4}}%
      \put(4988,375){\makebox(0,0){\strut{} 6}}%
      \put(5196,375){\makebox(0,0){\strut{} 8}}%
      \put(5403,375){\makebox(0,0){\strut{} 10}}%
      \put(5611,375){\makebox(0,0){\strut{} 12}}%
      \put(3683,1338){\rotatebox{-270}{\makebox(0,0){\strut{}}}}%
      \put(4988,89){\makebox(0,0){\strut{}}}%
    }%
    \gplgaddtomacro\gplfronttext{%
    }%
    \gplbacktext
    \put(0,0){\includegraphics{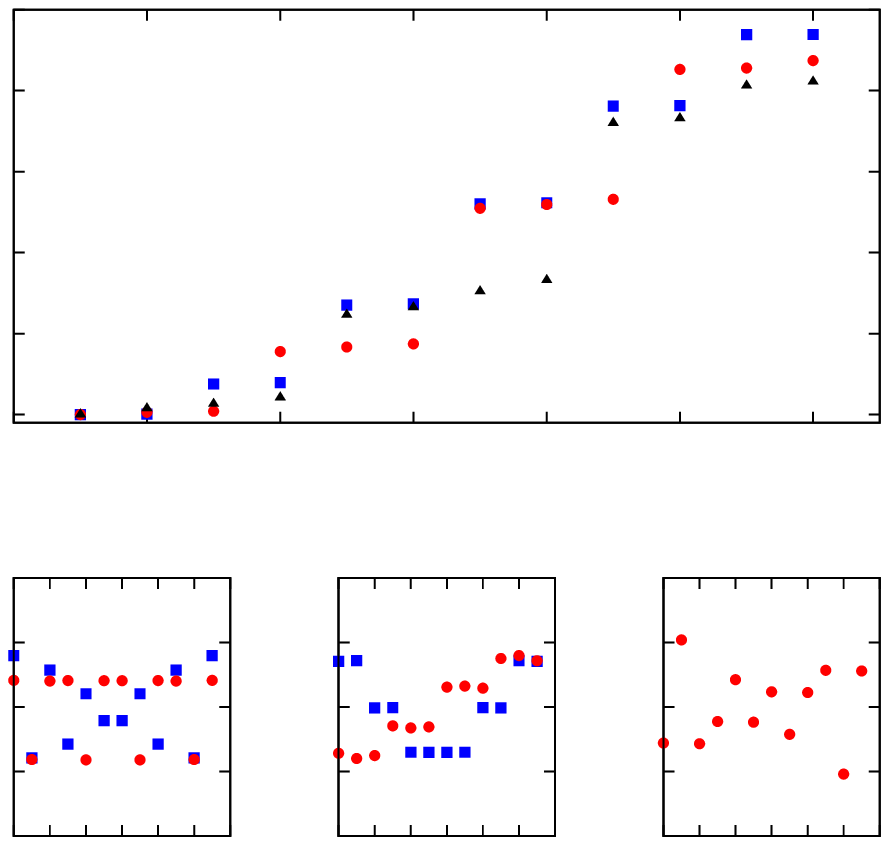}}%
    \gplfronttext
  \end{picture}%
\endgroup
\caption{(top) The spectrum of squared frequencies for $12$ dipoles,
  with two (blue square ), three (red circle) and four (black
  triangle) dipoles per revolution.  (bottom) The amplitudes of the
  normal modes for the lowest nearly degenerate eigenvalues. Left is
  eigenmode number 3 (blue square) and 4 (red circle), repectively for
  two and three dipoles per windings. Middle is eigenmode number 4
  (blue square) and 5 (red circle), repectively for two and three
  dipoles per windings. Right is eigenmode number 6 (red circle) for
  three dipoles per windings.  }
\label{fig:e12_234}
\end{figure*}

\subsection{Small systems}
The long-distance interaction is responsible for many properties.
In particular, it provides the attraction between particles on windings far
apart leading to collapse of the system for sufficiently large $N$.
By the same token small $N$ leads to stable systems with equidistant
spacing of the dipoles for moderate values of $h/R$, see
fig.~\ref{fig:Ncrit}.  In addition small $N$ allow faster
computations.  We can exploit this fact to study generic properties
like structures of low-lying normal modes.

We choose $N=12$ as this number of dipoles is divisible by both 2,3
and 4 dipoles per winding.  This match is expected to provide more
binding and very clean vibrational normal modes.  We show the lowest
frequencies in fig.~\ref{fig:e12_234} for $2,3$ and $4$ dipoles per
winding.  It is striking that the spectra are almost two-, three- and
four-fold degenerate.  This is caused by the commensurability to the
number of particles per winding and implies that the eigenvalues
appear either pairwise, as triples or as quadruplets.  The
degeneracies are getting less pronounced as the frequency increase.

These properties are signals of the chain structures discussed in the
previous subsections.  The computed degenerate normal modes correspond
for two dipoles per winding to the two chains moving either exactly in
phase preserving the same particle distances, or chains moving against
each other exactly in opposite phase.  In general, we characterize
each of the normal mode oscillations by a set of relative amplitudes,
where each describes the maximum deviation from equilibrium along the
helix of the corresponding dipole.  The amplitudes of these
oscillations are also shown in fig.~\ref{fig:e12_234}.  The same
degeneracies at higher frequencies correspond to the same vibrational
relation between chains but now with additional nodes along the
chains.

For two particles per winding the amplitudes for eigenmode number 3
(bottom left in fig.~\ref{fig:e12_234}) show alternating direction for
neighboring dipoles. This means that the two chains oscillate against
each other.  For three particles per winding the eigenmodes number
$4,5,6$ are shown from left to right in bottom panel of
fig.~\ref{fig:e12_234}.  Left show two chains moving against the third
as seen by two almost equal neighboring amplitudes and one amplitude
of about twice the size in opposite direction.  Similarly, in the
middle are the three chains in phase as the three neighboring
amplitudes are equal. On the right we see again two chains moving
against the third. 

For four particles per winding (not shown on figures) the four
normal modes are similarly related to chains moving in phase, or pairs
moving against each other, or triples against a single chain.  The
degeneracies and the normal mode structure are effects of the
next-nearest neighbour interaction, and the small difference within
the almost degenerate frequencies are due to small nearest-neighbour
corrections to the vibration-dominating next-nearest interaction.

\begin{figure*}
\centering
\begingroup
  \makeatletter
  \providecommand\color[2][]{%
    \GenericError{(gnuplot) \space\space\space\@spaces}{%
      Package color not loaded in conjunction with
      terminal option `colourtext'%
    }{See the gnuplot documentation for explanation.%
    }{Either use 'blacktext' in gnuplot or load the package
      color.sty in LaTeX.}%
    \renewcommand\color[2][]{}%
  }%
  \providecommand\includegraphics[2][]{%
    \GenericError{(gnuplot) \space\space\space\@spaces}{%
      Package graphicx or graphics not loaded%
    }{See the gnuplot documentation for explanation.%
    }{The gnuplot epslatex terminal needs graphicx.sty or graphics.sty.}%
    \renewcommand\includegraphics[2][]{}%
  }%
  \providecommand\rotatebox[2]{#2}%
  \@ifundefined{ifGPcolor}{%
    \newif\ifGPcolor
    \GPcolortrue
  }{}%
  \@ifundefined{ifGPblacktext}{%
    \newif\ifGPblacktext
    \GPblacktextfalse
  }{}%
  \let\gplgaddtomacro\g@addto@macro
  \gdef\gplbacktext{}%
  \gdef\gplfronttext{}%
  \makeatother
  \ifGPblacktext
    \def\colorrgb#1{}%
    \def\colorgray#1{}%
  \else
    \ifGPcolor
      \def\colorrgb#1{\color[rgb]{#1}}%
      \def\colorgray#1{\color[gray]{#1}}%
      \expandafter\def\csname LTw\endcsname{\color{white}}%
      \expandafter\def\csname LTb\endcsname{\color{black}}%
      \expandafter\def\csname LTa\endcsname{\color{black}}%
      \expandafter\def\csname LT0\endcsname{\color[rgb]{1,0,0}}%
      \expandafter\def\csname LT1\endcsname{\color[rgb]{0,1,0}}%
      \expandafter\def\csname LT2\endcsname{\color[rgb]{0,0,1}}%
      \expandafter\def\csname LT3\endcsname{\color[rgb]{1,0,1}}%
      \expandafter\def\csname LT4\endcsname{\color[rgb]{0,1,1}}%
      \expandafter\def\csname LT5\endcsname{\color[rgb]{1,1,0}}%
      \expandafter\def\csname LT6\endcsname{\color[rgb]{0,0,0}}%
      \expandafter\def\csname LT7\endcsname{\color[rgb]{1,0.3,0}}%
      \expandafter\def\csname LT8\endcsname{\color[rgb]{0.5,0.5,0.5}}%
    \else
      \def\colorrgb#1{\color{black}}%
      \def\colorgray#1{\color[gray]{#1}}%
      \expandafter\def\csname LTw\endcsname{\color{white}}%
      \expandafter\def\csname LTb\endcsname{\color{black}}%
      \expandafter\def\csname LTa\endcsname{\color{black}}%
      \expandafter\def\csname LT0\endcsname{\color{black}}%
      \expandafter\def\csname LT1\endcsname{\color{black}}%
      \expandafter\def\csname LT2\endcsname{\color{black}}%
      \expandafter\def\csname LT3\endcsname{\color{black}}%
      \expandafter\def\csname LT4\endcsname{\color{black}}%
      \expandafter\def\csname LT5\endcsname{\color{black}}%
      \expandafter\def\csname LT6\endcsname{\color{black}}%
      \expandafter\def\csname LT7\endcsname{\color{black}}%
      \expandafter\def\csname LT8\endcsname{\color{black}}%
    \fi
  \fi
  \setlength{\unitlength}{0.0500bp}%
  \begin{picture}(6236.00,5952.00)%
    \gplgaddtomacro\gplbacktext{%
      \csname LTb\endcsname%
      \put(491,3023){\makebox(0,0)[r]{\strut{} 0}}%
      \put(491,3489){\makebox(0,0)[r]{\strut{} 10}}%
      \put(491,3956){\makebox(0,0)[r]{\strut{} 20}}%
      \put(491,4422){\makebox(0,0)[r]{\strut{} 30}}%
      \put(491,4889){\makebox(0,0)[r]{\strut{} 40}}%
      \put(491,5355){\makebox(0,0)[r]{\strut{} 50}}%
      \put(623,2756){\makebox(0,0){\strut{} 0}}%
      \put(1336,2756){\makebox(0,0){\strut{} 2}}%
      \put(2048,2756){\makebox(0,0){\strut{} 4}}%
      \put(2761,2756){\makebox(0,0){\strut{} 6}}%
      \put(3473,2756){\makebox(0,0){\strut{} 8}}%
      \put(4186,2756){\makebox(0,0){\strut{} 10}}%
      \put(4898,2756){\makebox(0,0){\strut{} 12}}%
      \put(5611,2756){\makebox(0,0){\strut{} 14}}%
      \put(-147,4165){\rotatebox{-270}{\makebox(0,0){\strut{}$\omega^2$}}}%
      \put(3117,2470){\makebox(0,0){\strut{}}}%
      \put(3117,5685){\makebox(0,0){\strut{}Eigen mode}}%
    }%
    \gplgaddtomacro\gplfronttext{%
    }%
    \gplgaddtomacro\gplbacktext{%
      \csname LTb\endcsname%
      \put(491,595){\makebox(0,0)[r]{\strut{}-1}}%
      \put(491,967){\makebox(0,0)[r]{\strut{}-0.5}}%
      \put(491,1339){\makebox(0,0)[r]{\strut{} 0}}%
      \put(491,1710){\makebox(0,0)[r]{\strut{} 0.5}}%
      \put(491,2082){\makebox(0,0)[r]{\strut{} 1}}%
      \put(623,375){\makebox(0,0){\strut{} 0}}%
      \put(801,375){\makebox(0,0){\strut{} 2}}%
      \put(979,375){\makebox(0,0){\strut{} 4}}%
      \put(1157,375){\makebox(0,0){\strut{} 6}}%
      \put(1336,375){\makebox(0,0){\strut{} 8}}%
      \put(1514,375){\makebox(0,0){\strut{} 10}}%
      \put(1692,375){\makebox(0,0){\strut{} 12}}%
      \put(1870,375){\makebox(0,0){\strut{} 14}}%
      \put(-279,1338){\rotatebox{-270}{\makebox(0,0){\strut{}Amplitude}}}%
      \put(1246,89){\makebox(0,0){\strut{}}}%
    }%
    \gplgaddtomacro\gplfronttext{%
    }%
    \gplgaddtomacro\gplbacktext{%
      \csname LTb\endcsname%
      \put(4233,595){\makebox(0,0)[r]{\strut{}-1}}%
      \put(4233,967){\makebox(0,0)[r]{\strut{}-0.5}}%
      \put(4233,1339){\makebox(0,0)[r]{\strut{} 0}}%
      \put(4233,1710){\makebox(0,0)[r]{\strut{} 0.5}}%
      \put(4233,2082){\makebox(0,0)[r]{\strut{} 1}}%
      \put(4365,375){\makebox(0,0){\strut{} 0}}%
      \put(4543,375){\makebox(0,0){\strut{} 2}}%
      \put(4721,375){\makebox(0,0){\strut{} 4}}%
      \put(4899,375){\makebox(0,0){\strut{} 6}}%
      \put(5077,375){\makebox(0,0){\strut{} 8}}%
      \put(5255,375){\makebox(0,0){\strut{} 10}}%
      \put(5433,375){\makebox(0,0){\strut{} 12}}%
      \put(5611,375){\makebox(0,0){\strut{} 14}}%
      \put(3683,1338){\rotatebox{-270}{\makebox(0,0){\strut{}}}}%
      \put(2988,45){\makebox(0,0){\strut{}Dipole number along the helix}}%
    }%
    \gplgaddtomacro\gplfronttext{%
    }%
    \gplbacktext
    \put(0,0){\includegraphics{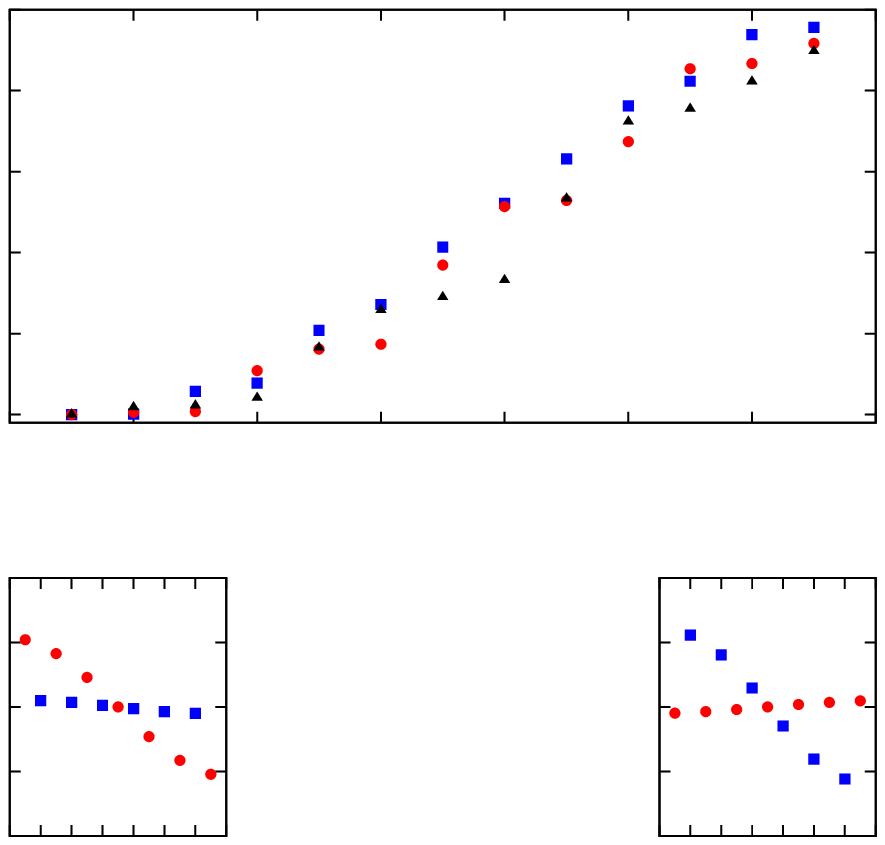}}%
    \gplfronttext
  \end{picture}%
\endgroup
\caption{(top) The spectrum of squared frequencies for $13$ dipoles,
  with two (blue square ), three (red circle) and four (black
  triangle) dipoles per revolution.  (bottom) The amplitudes of the
  normal modes for the lowest nearly degenerate eigenvalues of two
  dipoles per winding. Left is eigenmode number 3 where the long chain
  (red circle) is moving while the short chain (blue square) remains
  at rest. Right is eigenmode number 4 where the cjhains exchange
  motion. }
\label{fig:e13_234}
\end{figure*}

These simple structures arise so purely because an integer number of
particles can precisely fit into one winding. If this is not the case
the degeneracies are broken and the spectra would move towards another
type of degeneracy.  A non-integer ratio between particle number and
particles per winding must destroy the picture of chain structures,
simply because identical chains cannot be formed.  However, the lowest
energy from the strongest attraction arises by adding the odd
particles at the end of the chains.  These features are illustrated by
the example of $N=13$ shown in fig.~\ref{fig:e13_234}, where top panel
exhibits  frequencies for $2,3$ or $4$ dipoles per winding with the
additional dipole placed at the end of one of the chains.

The degeneracies compared to N = 12 are now lifted
although traces can be detected.  However, the modes exhibited in
fig.~\ref{fig:e12_234} still come in pairs for two dipoles per
winding, even though their frequencies differ.  The same features of
normal mode similarity between systems with $N=12$ and $13$ are also
seen for the energies of three and four particles per winding.  The
amplitudes for two modes of two dipoles per winding at the bottom of
fig.~\ref{fig:e12_234} show that one of the chains essentially is not
moving while the other is undergoing an odd parity oscillation.  The
interpretation is in general that the slightly different chains still
vibrate against each other.  These boundary layer effects would
decrease with increasing total number of particles.

\section{Speed of Sound}
The speed of sound in a crystal is a fundamental property which plays
an important role in many physical processes.  The sound is mediated
by particle motion and normal modes carry the necessary information
about the correlated small amplitude vibrations which form the sound
waves.  The length of our one-dimensional regular crystal is $N R
(\Delta \phi)_0$, where $(\Delta \phi)_0 = \phi_{i+1} - \phi_i$ is the
distance between equally spaced neighbouring dipoles along the helix.
The corresponding wave numbers are $k_n = 2\pi n/(N R (\Delta \phi)_0)
$ where $n$ is an integer.  The speed of sound is then given by $c =
\omega_n/k_n $ in the limit of small $k_n$ when $\omega_n$ is the n'th
normal mode frequency.  Since small $k_n$ requires large $N$ we have
to find normal modes for large systems.

\subsection{Simple models}
The restoring force is in general obtained from the energy in
Eq.~(\ref{eq:deriv2}) for arbitrary, independent displacements of the
individual dipoles.  The stable configurations are equidistantly
distributed dipoles along the helix for a modest density per
revolution and perhaps also only for a finite total number of
particles.  For this structure we search for small amplitude
vibrations around equilibrium which allows sufficiently accurate
energy calculation by using second order as in Eq.~(\ref{eq:deriv2}).
Large $N$ is expected to eliminate dependence on specific choices of
end-point configurations.  We shall explore consequences of these
approximations while staying within the validity ranges found in the
previous sections.

The equal spacing condition for a very long sequence of dipoles has
several consequences we must take into account.  First, the same
translation of all dipoles does not change any of the relative
distances, and the total energy must remain unchanged as well. Second,
all two-body properties only depend on the relative distance between
the particles.  In particular, the value of $K_{ij}$ is then the same
as $K_{k+i,k+j}$ for any value of $k$.  The double sum in
Eq.~(\ref{eq:deriv2}) can then be performed in different order where
the summation indices are along constant sum, $k$, and constant
difference, $l$. However, only relative distances produce any energy
variation.  Let us measure deviations from equilibrium positions for
the individual variables, that is using the coordinates
$\tilde{\phi}_i=\phi_i-\phi_i^{(0)}$. In addition, we assume that the
energy to second order in distances from equilibrium is independent of
which particle is used as the reference position.  These assumptions
lead to
\begin{equation} \label{eq:deriv2a}
 E-E_0= \sum_{l>0}\frac{1}{2} K_{k+l,k-l}  
 \left(\tilde{\phi}_{k+l}-\tilde{\phi}_k\right)\left(\tilde{\phi}_{k-l} 
 - \tilde{\phi}_k\right) \; ,
\end{equation}
where we measure relative to particle $k$.  Eq.(\ref{eq:deriv2a}) is
then independent of $k$ and the curvature only depends on $l$.

The classical force on particle $n$ is minus the gradient of the
potential energy given in Eq.~(\ref{eq:deriv2a}) which according to
Newton's second law equals mass times acceleration given as the second
time derivative.  The equation of motion becomes
\begin{eqnarray}  \label{eq:newton1}
  m \frac{\partial^2\tilde{\phi}_k}{\partial t^2} 
   =  \frac{1}{2} \sum_{l>0} K_l 
 (\tilde{\phi}_{k+l} + \tilde{\phi}_{k-l}-2\tilde{\phi}_{k})   \; ,
\end{eqnarray}
where we used the definitions $ K_l = K_{k+l,k-l}$.  We now search for
periodic motion, $\tilde{\phi}_j = A_j \cos(\omega t + a_j)$, around
the $j$ equilibrium positions.  The time dependence cancel on both
sides of Eq.~(\ref{eq:newton1}) if and only if the amplitudes are
identical, that is $A_j= A_0$, and simultaneously the average phase
$(a_{k-l} + a_{k+l})/2$ equals $a_{k}$ for all $l$.  The latter
condition immediately implies a linear dependence of the phases, that
is $a_{j} = b_0 + j b_1$.  With these definitions we arrive at 
the frequency given by  
\begin{eqnarray} 
  m \omega^2 & =&  \sum_{l>0} K_{l} (1 - \cos(lb_1))  \nonumber \\
  &=&  2 \sum_{l>0} K_{l} \sin^2(lb_1/2)  \; . \label{eq:newtonfreq}
\end{eqnarray}
These frequencies with the corresponding eigen phases given by $b_1$
characterize the $N$ independent periodic solutions. The spectrum is
continuous as any value of $b_1$ between $0$ and $2\pi$ provide a
frequency through Eq.~(\ref{eq:newtonfreq}). An appropriate
discretization can for example be found by assuming the same phase,
apart from a multiplum of $2\pi$, for the end-point dipoles.  This means
that $a_0 = a_N + n 2 \pi$ which is equivalent to $b_1 = 2 \pi n/N$ for
any integer value $n$.  This could be obtained by bending the helix to
let the two end-points meet.  Now the $N$ independent solutions,
$\omega_n$, result from Eq.~(\ref{eq:newtonfreq}).

The size of $K_l$ decreases strongly with $l$, since the overall
energy scale is obtained from a cubic decrease with distance.  The
largest $K_l$ is then usually by far the lowest $l$, that is $l=1$.
By using only lowest order in an expansion of
Eq.~(\ref{eq:newtonfreq}), we find the lowest frequencies to be
approximately given by
\begin{eqnarray} \label{eq:newtonfreq2}
 \omega_n \approx \frac{2\pi n}{N} \sqrt{ \sum_{l>0} l^2 K_{l}/m} \;. 
\end{eqnarray}
The speed of sound, $c$, is then obtained as the ratio between
$\omega_n$ and the wave number $k_n$, where the latter is found by
equating the total length of the helix with the wave length. Thus,
$R N (\Delta \phi)_0 = 2\pi n /k_n$, where $(\Delta \phi)_0$ is
the angular distance between the equidistantly placed dipoles.  We
then finally get
\begin{eqnarray} \label{eq:sspeed}
c = \frac{\omega_n }{k_n} \approx R (\Delta \phi)_0 
\sqrt{ \sum_{l>0} l^2 K_{l}/m}  \; ,
\end{eqnarray}
which is the result for a periodic structure of particles located on a
ring \cite{Ashcroft}.  Thus we get the same $c$ independent of which
of the lowest modes of $n=1,2,3$ we used.  Furthermore, $c$ is also
roughly independent of the number of dipoles per winding.  This is
seen from Eq.~(\ref{eq:newtonfreq}) when the $K_l$-summation is
dominated by the first in-chain contribution for $l=l_0$ giving $c = R
(\Delta \phi)_0 l_0 \sqrt{K_{l_0}/m}$.  We then obtain the same $c$,
since both $K_{l_0}$ and $(\Delta \phi)_0 l_0$ are independent of
$l_0$ or equivalently dipole density.

\begin{figure}
\begingroup
  \makeatletter
  \providecommand\color[2][]{%
    \GenericError{(gnuplot) \space\space\space\@spaces}{%
      Package color not loaded in conjunction with
      terminal option `colourtext'%
    }{See the gnuplot documentation for explanation.%
    }{Either use 'blacktext' in gnuplot or load the package
      color.sty in LaTeX.}%
    \renewcommand\color[2][]{}%
  }%
  \providecommand\includegraphics[2][]{%
    \GenericError{(gnuplot) \space\space\space\@spaces}{%
      Package graphicx or graphics not loaded%
    }{See the gnuplot documentation for explanation.%
    }{The gnuplot epslatex terminal needs graphicx.sty or graphics.sty.}%
    \renewcommand\includegraphics[2][]{}%
  }%
  \providecommand\rotatebox[2]{#2}%
  \@ifundefined{ifGPcolor}{%
    \newif\ifGPcolor
    \GPcolortrue
  }{}%
  \@ifundefined{ifGPblacktext}{%
    \newif\ifGPblacktext
    \GPblacktextfalse
  }{}%
  \let\gplgaddtomacro\g@addto@macro
  \gdef\gplbacktext{}%
  \gdef\gplfronttext{}%
  \makeatother
  \ifGPblacktext
    \def\colorrgb#1{}%
    \def\colorgray#1{}%
  \else
    \ifGPcolor
      \def\colorrgb#1{\color[rgb]{#1}}%
      \def\colorgray#1{\color[gray]{#1}}%
      \expandafter\def\csname LTw\endcsname{\color{white}}%
      \expandafter\def\csname LTb\endcsname{\color{black}}%
      \expandafter\def\csname LTa\endcsname{\color{black}}%
      \expandafter\def\csname LT0\endcsname{\color[rgb]{1,0,0}}%
      \expandafter\def\csname LT1\endcsname{\color[rgb]{0,1,0}}%
      \expandafter\def\csname LT2\endcsname{\color[rgb]{0,0,1}}%
      \expandafter\def\csname LT3\endcsname{\color[rgb]{1,0,1}}%
      \expandafter\def\csname LT4\endcsname{\color[rgb]{0,1,1}}%
      \expandafter\def\csname LT5\endcsname{\color[rgb]{1,1,0}}%
      \expandafter\def\csname LT6\endcsname{\color[rgb]{0,0,0}}%
      \expandafter\def\csname LT7\endcsname{\color[rgb]{1,0.3,0}}%
      \expandafter\def\csname LT8\endcsname{\color[rgb]{0.5,0.5,0.5}}%
    \else
      \def\colorrgb#1{\color{black}}%
      \def\colorgray#1{\color[gray]{#1}}%
      \expandafter\def\csname LTw\endcsname{\color{white}}%
      \expandafter\def\csname LTb\endcsname{\color{black}}%
      \expandafter\def\csname LTa\endcsname{\color{black}}%
      \expandafter\def\csname LT0\endcsname{\color{black}}%
      \expandafter\def\csname LT1\endcsname{\color{black}}%
      \expandafter\def\csname LT2\endcsname{\color{black}}%
      \expandafter\def\csname LT3\endcsname{\color{black}}%
      \expandafter\def\csname LT4\endcsname{\color{black}}%
      \expandafter\def\csname LT5\endcsname{\color{black}}%
      \expandafter\def\csname LT6\endcsname{\color{black}}%
      \expandafter\def\csname LT7\endcsname{\color{black}}%
      \expandafter\def\csname LT8\endcsname{\color{black}}%
    \fi
  \fi
  \setlength{\unitlength}{0.0500bp}%
  \begin{picture}(4676.00,2834.00)%
    \gplgaddtomacro\gplbacktext{%
      \csname LTb\endcsname%
      \put(814,440){\makebox(0,0)[r]{\strut{} 0}}%
      \put(814,795){\makebox(0,0)[r]{\strut{} 10}}%
      \put(814,1150){\makebox(0,0)[r]{\strut{} 20}}%
      \put(814,1505){\makebox(0,0)[r]{\strut{} 30}}%
      \put(814,1859){\makebox(0,0)[r]{\strut{} 40}}%
      \put(814,2214){\makebox(0,0)[r]{\strut{} 50}}%
      \put(814,2569){\makebox(0,0)[r]{\strut{} 60}}%
      \put(946,220){\makebox(0,0){\strut{} 0}}%
      \put(1613,220){\makebox(0,0){\strut{} 0.2}}%
      \put(2279,220){\makebox(0,0){\strut{} 0.4}}%
      \put(2946,220){\makebox(0,0){\strut{} 0.6}}%
      \put(3612,220){\makebox(0,0){\strut{} 0.8}}%
      \put(4279,220){\makebox(0,0){\strut{} 1}}%
      \put(176,1504){\rotatebox{-270}{\makebox(0,0){\strut{}$\omega^2$}}}%
      \put(2612,-66){\makebox(0,0){\strut{}}}%
    }%
    \gplgaddtomacro\gplfronttext{%
    }%
    \gplbacktext
    \put(0,0){\includegraphics{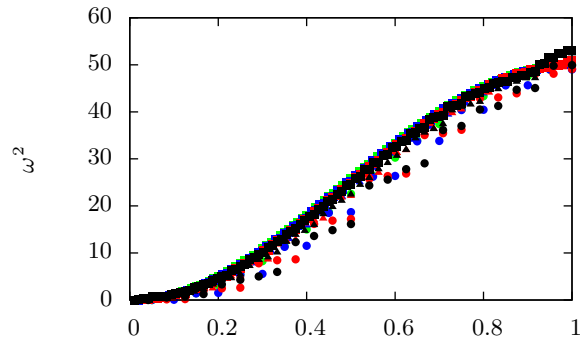}}%
    \gplfronttext
  \end{picture}%
\endgroup
\caption{The squared eigenfrequencies with $h/R=1$ for $N = 20$ (circle), $50$
  (triangle), and $100$ (square) as functions of their number in the
  ordered sets of eigenvalues divided by the total number of dipoles.  
  The dipoles per revolution are $1$ (green),
  $2$ (blue), $3$ (red), and $4$ (black).
 \label{fig:freq1}}
\end{figure}

\subsection{Numerical results}
We calculate all $N$ eigenfrequencies, $\omega_n$, as functions of
$N$ for $1,2,3,4$ dipoles per revolution in the equidistant stable
configurations.  We first show the ordered frequencies for
$N=20,50,100$ and $h=R$ in fig.~\ref{fig:freq1}.  The $x$-axis is
scaled with the different number of dipoles.  The details are not
meant to be visible on this figure, where the striking features are
that all eigenfrequencies, after scaling, essentially follow the same
curve varying from zero to a maximum value of about $50$ for the
largest frequency.

This behaviour is evident from Eq.~(\ref{eq:newtonfreq}) when the
discretization of $b_1$ is considered to vary from $0$ to $2\pi$.
These two end-point values of $b_1$ result in the maximum value of the
cosine function, and a limiting largest value obtained as the sum of
all the curvatures.  This sum is essentially independent of $N$, since
$K_l$ decreases very fast with the distance between the dipoles as
parametrized by the index $l$.  Thus, only the smallest values of $l$
then contribute in the summation.  Values of $b_1$ around $\pi$
produce alternating signs of the cosine function and give the
small frequencies.

\begin{figure}
\begingroup
  \makeatletter
  \providecommand\color[2][]{%
    \GenericError{(gnuplot) \space\space\space\@spaces}{%
      Package color not loaded in conjunction with
      terminal option `colourtext'%
    }{See the gnuplot documentation for explanation.%
    }{Either use 'blacktext' in gnuplot or load the package
      color.sty in LaTeX.}%
    \renewcommand\color[2][]{}%
  }%
  \providecommand\includegraphics[2][]{%
    \GenericError{(gnuplot) \space\space\space\@spaces}{%
      Package graphicx or graphics not loaded%
    }{See the gnuplot documentation for explanation.%
    }{The gnuplot epslatex terminal needs graphicx.sty or graphics.sty.}%
    \renewcommand\includegraphics[2][]{}%
  }%
  \providecommand\rotatebox[2]{#2}%
  \@ifundefined{ifGPcolor}{%
    \newif\ifGPcolor
    \GPcolortrue
  }{}%
  \@ifundefined{ifGPblacktext}{%
    \newif\ifGPblacktext
    \GPblacktextfalse
  }{}%
  \let\gplgaddtomacro\g@addto@macro
  \gdef\gplbacktext{}%
  \gdef\gplfronttext{}%
  \makeatother
  \ifGPblacktext
    \def\colorrgb#1{}%
    \def\colorgray#1{}%
  \else
    \ifGPcolor
      \def\colorrgb#1{\color[rgb]{#1}}%
      \def\colorgray#1{\color[gray]{#1}}%
      \expandafter\def\csname LTw\endcsname{\color{white}}%
      \expandafter\def\csname LTb\endcsname{\color{black}}%
      \expandafter\def\csname LTa\endcsname{\color{black}}%
      \expandafter\def\csname LT0\endcsname{\color[rgb]{1,0,0}}%
      \expandafter\def\csname LT1\endcsname{\color[rgb]{0,1,0}}%
      \expandafter\def\csname LT2\endcsname{\color[rgb]{0,0,1}}%
      \expandafter\def\csname LT3\endcsname{\color[rgb]{1,0,1}}%
      \expandafter\def\csname LT4\endcsname{\color[rgb]{0,1,1}}%
      \expandafter\def\csname LT5\endcsname{\color[rgb]{1,1,0}}%
      \expandafter\def\csname LT6\endcsname{\color[rgb]{0,0,0}}%
      \expandafter\def\csname LT7\endcsname{\color[rgb]{1,0.3,0}}%
      \expandafter\def\csname LT8\endcsname{\color[rgb]{0.5,0.5,0.5}}%
    \else
      \def\colorrgb#1{\color{black}}%
      \def\colorgray#1{\color[gray]{#1}}%
      \expandafter\def\csname LTw\endcsname{\color{white}}%
      \expandafter\def\csname LTb\endcsname{\color{black}}%
      \expandafter\def\csname LTa\endcsname{\color{black}}%
      \expandafter\def\csname LT0\endcsname{\color{black}}%
      \expandafter\def\csname LT1\endcsname{\color{black}}%
      \expandafter\def\csname LT2\endcsname{\color{black}}%
      \expandafter\def\csname LT3\endcsname{\color{black}}%
      \expandafter\def\csname LT4\endcsname{\color{black}}%
      \expandafter\def\csname LT5\endcsname{\color{black}}%
      \expandafter\def\csname LT6\endcsname{\color{black}}%
      \expandafter\def\csname LT7\endcsname{\color{black}}%
      \expandafter\def\csname LT8\endcsname{\color{black}}%
    \fi
  \fi
  \setlength{\unitlength}{0.0500bp}%
  \begin{picture}(4676.00,2834.00)%
    \gplgaddtomacro\gplbacktext{%
      \csname LTb\endcsname%
      \put(814,704){\makebox(0,0)[r]{\strut{} 0}}%
      \put(814,1077){\makebox(0,0)[r]{\strut{} 10}}%
      \put(814,1450){\makebox(0,0)[r]{\strut{} 20}}%
      \put(814,1823){\makebox(0,0)[r]{\strut{} 30}}%
      \put(814,2196){\makebox(0,0)[r]{\strut{} 40}}%
      \put(814,2569){\makebox(0,0)[r]{\strut{} 50}}%
      \put(946,484){\makebox(0,0){\strut{} 0}}%
      \put(1779,484){\makebox(0,0){\strut{} 500}}%
      \put(2613,484){\makebox(0,0){\strut{} 1000}}%
      \put(3446,484){\makebox(0,0){\strut{} 1500}}%
      \put(4279,484){\makebox(0,0){\strut{} 2000}}%
      \put(176,1636){\rotatebox{-270}{\makebox(0,0){\strut{}$\frac{\omega_iN}{i}$}}}%
      \put(2612,154){\makebox(0,0){\strut{}$N$}}%
    }%
    \gplgaddtomacro\gplfronttext{%
    }%
    \gplbacktext
    \put(0,0){\includegraphics{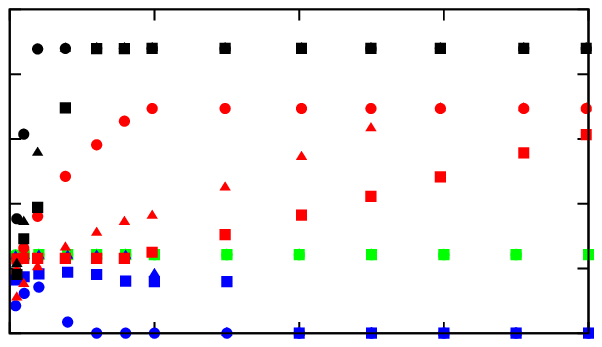}}%
    \gplfronttext
  \end{picture}%
\endgroup
\caption{The lowest three eigenfrequencies multiplied by $N$ and
  divided by respectively $1$ (circle), $2$ (triangle), and $3$
  (square) exhibited as functions of $N$.  We choose $h=R$ and the
  number of dipoles per revolution to be $1$ (green), $2$ (blue), $3$
  (red), and $4$ (black).
 \label{fig:freq2}}
\end{figure}

We now focus on the lowest frequencies which according to
Eq.~(\ref{eq:newtonfreq}) are expected to vanish as $n/N$ for
increasing $N$.  We therefore show the frequencies multiplied by $N/n$
in fig.~\ref{fig:freq2} as functions of $N$ for the previous case of a
few dipoles per revolution.  The linear dependence is quickly reached
for most of the stable systems. This arises from the allowed lowest
order substitution of the argument in the sine function of
Eq.~(\ref{eq:newtonfreq}).  Here $2 \pi n /N \ll 1$ or equivalently
$N$ must considerably exceed $6 n = 6,12,18$, which easily should be
fulfilled for our cases.  Also the increase proportional to $n$ is
seen in all cases.

An exception in fig.~\ref{fig:freq2} is seen for two dipoles per
revolution where the small $N$ increase is followed by a maximum and a
subsequent decrease. This behaviour reflects the transition towards
instability of the configuration of equidistant spacing.  Another
exception appears for three particles per revolution.  The constant
values approached by the second and third lowest frequencies occur
apparently for much larger values of $N$, that is about $500$ and
$1500$.  The latter anomaly can be explained by starting with very
small $N$ where we know that a near three-fold degeneracy appear.
Increasing $N$ the two lowest frequencies increase, whereas the third
stabilizes and remain at this constant value when $N$ passes $500$.
The other two, initially lowest, frequencies continue to increase
until about $1500$ where a new stabilization takes place due to the
crossing of an initially much higher lying mode.

The stability of $N\omega_n$ for large $N$ reflect the appearance of
acoustic modes which, as we shall discuss now, corresponds to in-chain
dipole oscillations.  The frequency crossings of these modes by
lower-lying modes for small $N$ reflect that out-of-chain oscillations
are more favorable for small chains.

The features discussed here are exemplified with $h=R$ results.
However, the general behaviour for other helix parameters can be
understood with the same ingredients.  The curvatures would change and
accordingly also the frequencies.  The only other differences are that
the stabilities depend rather strongly on $h/R$ and the large
$N$-limit and the resulting linear dependence on $1/N$ may sometimes
not be reached for the equidistant configurations.

\begin{figure}
\begingroup
  \makeatletter
  \providecommand\color[2][]{%
    \GenericError{(gnuplot) \space\space\space\@spaces}{%
      Package color not loaded in conjunction with
      terminal option `colourtext'%
    }{See the gnuplot documentation for explanation.%
    }{Either use 'blacktext' in gnuplot or load the package
      color.sty in LaTeX.}%
    \renewcommand\color[2][]{}%
  }%
  \providecommand\includegraphics[2][]{%
    \GenericError{(gnuplot) \space\space\space\@spaces}{%
      Package graphicx or graphics not loaded%
    }{See the gnuplot documentation for explanation.%
    }{The gnuplot epslatex terminal needs graphicx.sty or graphics.sty.}%
    \renewcommand\includegraphics[2][]{}%
  }%
  \providecommand\rotatebox[2]{#2}%
  \@ifundefined{ifGPcolor}{%
    \newif\ifGPcolor
    \GPcolortrue
  }{}%
  \@ifundefined{ifGPblacktext}{%
    \newif\ifGPblacktext
    \GPblacktextfalse
  }{}%
  \let\gplgaddtomacro\g@addto@macro
  \gdef\gplbacktext{}%
  \gdef\gplfronttext{}%
  \makeatother
  \ifGPblacktext
    \def\colorrgb#1{}%
    \def\colorgray#1{}%
  \else
    \ifGPcolor
      \def\colorrgb#1{\color[rgb]{#1}}%
      \def\colorgray#1{\color[gray]{#1}}%
      \expandafter\def\csname LTw\endcsname{\color{white}}%
      \expandafter\def\csname LTb\endcsname{\color{black}}%
      \expandafter\def\csname LTa\endcsname{\color{black}}%
      \expandafter\def\csname LT0\endcsname{\color[rgb]{1,0,0}}%
      \expandafter\def\csname LT1\endcsname{\color[rgb]{0,1,0}}%
      \expandafter\def\csname LT2\endcsname{\color[rgb]{0,0,1}}%
      \expandafter\def\csname LT3\endcsname{\color[rgb]{1,0,1}}%
      \expandafter\def\csname LT4\endcsname{\color[rgb]{0,1,1}}%
      \expandafter\def\csname LT5\endcsname{\color[rgb]{1,1,0}}%
      \expandafter\def\csname LT6\endcsname{\color[rgb]{0,0,0}}%
      \expandafter\def\csname LT7\endcsname{\color[rgb]{1,0.3,0}}%
      \expandafter\def\csname LT8\endcsname{\color[rgb]{0.5,0.5,0.5}}%
    \else
      \def\colorrgb#1{\color{black}}%
      \def\colorgray#1{\color[gray]{#1}}%
      \expandafter\def\csname LTw\endcsname{\color{white}}%
      \expandafter\def\csname LTb\endcsname{\color{black}}%
      \expandafter\def\csname LTa\endcsname{\color{black}}%
      \expandafter\def\csname LT0\endcsname{\color{black}}%
      \expandafter\def\csname LT1\endcsname{\color{black}}%
      \expandafter\def\csname LT2\endcsname{\color{black}}%
      \expandafter\def\csname LT3\endcsname{\color{black}}%
      \expandafter\def\csname LT4\endcsname{\color{black}}%
      \expandafter\def\csname LT5\endcsname{\color{black}}%
      \expandafter\def\csname LT6\endcsname{\color{black}}%
      \expandafter\def\csname LT7\endcsname{\color{black}}%
      \expandafter\def\csname LT8\endcsname{\color{black}}%
    \fi
  \fi
  \setlength{\unitlength}{0.0500bp}%
  \begin{picture}(4676.00,2834.00)%
    \gplgaddtomacro\gplbacktext{%
      \csname LTb\endcsname%
      \put(946,704){\makebox(0,0)[r]{\strut{} 0}}%
      \put(946,1015){\makebox(0,0)[r]{\strut{} 20}}%
      \put(946,1326){\makebox(0,0)[r]{\strut{} 40}}%
      \put(946,1637){\makebox(0,0)[r]{\strut{} 60}}%
      \put(946,1947){\makebox(0,0)[r]{\strut{} 80}}%
      \put(946,2258){\makebox(0,0)[r]{\strut{} 100}}%
      \put(946,2569){\makebox(0,0)[r]{\strut{} 120}}%
      \put(1078,484){\makebox(0,0){\strut{} 0.2}}%
      \put(1535,484){\makebox(0,0){\strut{} 0.4}}%
      \put(1993,484){\makebox(0,0){\strut{} 0.6}}%
      \put(2450,484){\makebox(0,0){\strut{} 0.8}}%
      \put(2907,484){\makebox(0,0){\strut{} 1}}%
      \put(3364,484){\makebox(0,0){\strut{} 1.2}}%
      \put(3822,484){\makebox(0,0){\strut{} 1.4}}%
      \put(4279,484){\makebox(0,0){\strut{} 1.6}}%
      \put(176,1636){\rotatebox{-270}{\makebox(0,0){\strut{}Speed of sound $c$}}}%
      \put(2678,154){\makebox(0,0){\strut{}$\frac{h}{R}$}}%
    }%
    \gplgaddtomacro\gplfronttext{%
    }%
    \gplbacktext
    \put(0,0){\includegraphics{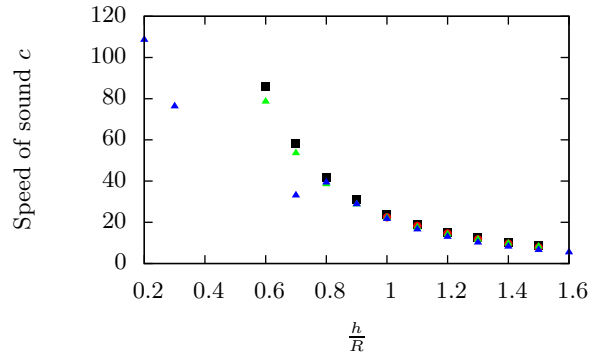}}%
    \gplfronttext
  \end{picture}%
\endgroup
\caption{The Speed of sound c obtained for $n=1$ as a function of the
  ratio $\frac{h}{R}$ for one dipole per revolution (black square),
  three dipoles per revolution (red circle), four dipoles per
  revolution (blue nabla) and a nearest-neighbour calculation for 1
  dipole (green triangle) \label{fig:cofhR}}
\end{figure}

We now proceed to calculate the speed of sound by dividing the
frequencies by the wave number as in eq.~(\ref{eq:sspeed}).  This
means multiplying the numbers in fig.~\ref{fig:freq2} with $R (\Delta
\phi)_0)/(2\pi n)$ where $n$ is 1, 2 and 3.  The results are shown in
fig.~\ref{fig:cofhR} for $n=1$ as function of $h/R$.  A smooth curve
for one dipole per revolution is found for a large range of helix
parameters. The speed of sound decreases roughly as $(R/h)^3$ with
increasing $h/R$. This is almost the same scaling as observed for the
energy, see fig.~\ref{fig:EofhR}. These results are calculated from
the lowest frequency but essentially the same curve appears by use of
the second and third frequencies corresponding to $n=2,3$.  The
proportionality in $n$ from Eq.~(\ref{eq:sspeed}) is cancelled by that
in $k_{n}$.

More dipoles per revolution also lead to series of points in
fig.~\ref{fig:EofhR}.  However, instability at large $N$ before
stabilization prevents calculation of the speed of sound and fewer
points can be obtained.  It is striking that all available points
appear essentially on the same curve as the one-chain result.  The
equidistant equilibrium spacing, $(\Delta \phi)_0$, decreases
with the number of dipoles per revolution. The factor of $l^2$
inside the square-root in Eq.~(\ref{eq:sspeed}) can compensate this
when a single value of $K_l$ dominates the speed of sound.
This happens in the chain picture. Here $K_l$ will be small 
for $l$ less than the number of chains, $l_0$. However, 
$K_{l_0}$ will be the term describing the interaction with the 
nearest-neighbour within a single chain. The $K_l$ terms 
for $l<l_0$ are interactions with particles in different 
chains. In summary, we have a chain dominated dynamics where
we can use an in-chain nearerst-neighbour approximation for 
any number of chains.
Thus, the sound waves
travel within each chain, and the perturbations of the waves within a
single chain from neighbouring chains are minimal.

Finally, we investigate the consequences of the instability of some of
the configurations.  For $h=R$ we know that the chains prefer specific
distances for sufficiently large $N$ as exemplified in
figs.~\ref{fig:EofChain} and \ref{fig:h06}.  We therefore calculate
the normal mode spectra for the non-equidistant equilibrium
configurations of two dipoles per winding.  The resulting points for
an average value, $(\Delta \phi)_0$, fall on the common curve in
fig.~\ref{fig:cofhR} found for the equidistant configurations.  This
again support the interpretation in terms of chain structures given
above.

\section{Discussion and Outlook}
The classical potential energy is calculated for a total large number,
$N$, of dipoles placed on a helix and uniformly distributed with a
density corresponding to $2,3$ and $4$ dipoles per winding.  The
dependence on the ratio of the two helix lengths, pitch-to-radius
($h/R$), is investigated.  The unit of energy is given by dipole
moment squared divided by the cubic power of the radius. These
parameters only enter as a scale without influence on the geometric
structures providing stability.  The one dimensional confinement
implies that the dipoles prefer to be located at specific angular
distances from each other along the helix.  This is due to the strong
attraction between head-to-tail configurations combined with equally
strong side-by-side repulsion.  The system collapses into strings of
infinite energy when $h>R \pi \sqrt{2}$, which therefore is a limit we
do not exceed.

The stable structures for relatively small densities can be
characterized as a number of chains formed by dipoles almost
head-to-tail along the helix. The stable chain structures are
equidistantly distributed when $h/R$ is larger than about 1 and $N$
is relatively small. Increasing $N$ for these $h/R$ values leads to
instability.  For decreasing $h/R$ stability is maintained for larger
$N$ but eventually lost for all densities.  The structures with 
lower energy that cause this instability in the case with 
a few dipoles per winding are clusterized states where chains
of dipoles cluster. These modes of
instability are investigated through the classical normal modes. 
An abrupt change of the chain structures occurs when $h/R$ is only
marginally varied. This is analogous to spontaneou symmetry breaking
of a symmetric minimum by conversion into a maximum and two asymmetric
minima. A main finding is that 
both initial and final structures can be characterized as
chains and thus the change is really in the local density due to 
clusterization.

By increasing the density the preferred configuration changes from one
to more chains. For densities correponding to less than five dipoles
per winding the equidistant spacing around the circles are broken for
a sufficiently extended helix.  The more chains imposed, the more
stability because the short-range repulsion between neighboring
dipoles on the circles lock each other in positions on average as far
away as possible.  This is classical crystal formation.

Detailed investigations are also carried out for smaller $N$ where we
find signals of preferred geometries in degeneracies, or lack of them,
in the normal mode frquenies.  For example $N=12$ allow very symmetric
structures of $2$, $3$ and $4$ chains with corresponding normal mode
degeneracies of $2$, $3$ and $4$.  The normal mode amplitudes reflect
preference for chains as inert units vibrating against each other.
The degeneracies are broken by addition of one more dipole at the end
of one chain.  However, the chains still prefer to vibrate in the same
way as inert entities. 
Finally we calculate the speed of sound whenever we find stable
structures for sufficiently large $N$.  We find the same behaviour of
the speed of sound as
function of $h/R$ for $2,3$ and $4$ chains in the system.  
This confirms the chain
picture as it shows that sound waves propagate along the chains.

In this paper we investigated a finite number of dipoles on a finite
helix. Boundary conditions corresponding to densities of a non-integer
average number of dipoles per winding were only discussed for a few
small systems.  For larger systems the optimum configurations would be
chains found by dividing the dipole number by the number of windings
and rounding up to the nearest interger. This leaves room for adding
the non-matching dipoles at the outer circle.  They would each prefer
to be located in continuation of one of the chains.  For very large
systems these end-point structures are not expected to influence the
bulk structures.

In future studies, we want to consider the quantum effects on the 
system both from the few-body and the many-body point of view. For
few-body quantum bound states, previous studies have considered 
dipolar particles in one-dimensional tubes and on rings \cite{deuret2010,klawunn2010,zollner2011a,zollner2011b,artem2013}.
Dipolar interactions can be addressed using harmonic approximations \cite{jeremy2011,jeremy2012,fil2014}, 
stochastic variational methods \cite{artem2012}, and exact diagonalization \cite{cremon2010,zinner2011,kristin2013}.
All of these methods should be adaptable to the helical geometry. 
For treatment of larger particle numbers, we imagine that a 
combination of the Luttinger approaches with dipolar particles
\cite{law2008,citro2007,citro2008,chang2009,huang2009,dalmonte2010,kollath2008,leche2012},
and matrix product states \cite{verstrate2008} or density matrix renormalization group methods \cite{white1992,scholl2005}
are possible approaches. Some recent examples of the numerical methods applied
to dipolar particles can be found in Refs.~\cite{arguelles2007,knap2012,ruhman2012,gammelmark2014}.

As discussed briefly in the introduction, we expect that the most 
straighforward experimental realization of helical geometries will 
be in systems where light is guided by an optical nanofiber with a 
diameter that is smaller than the wave length of the light such that
an evanescent wave builds that can be used as an atomic trap 
This has been done by the Rauschenbeutel group in Vienna \cite{sague2008,vetsch2010,dawkins2011}. 
The same group has suggested the creation of helical traps in a
recent paper \cite{reitz2012}. At the moment these proposals 
involve trapping of non-polar Rubidium atoms and thus one would 
need to extend the technique to either polar molecules or to 
atoms with large intrinsic magnetic dipolar moments. Given the 
added complications in producing cold polar molecules, it may be
more easy to use magnetic dipoles.

Our study here concerns a single helix. An obvious extension is to the double
helix DNA geometry proposed in Ref.~\cite{reitz2012}. Based on the current 
finding one might expect that this would merely be a doubled version of the 
single helix ground states. However, due to the long-range interactions 
one could imagine that more involved states would be allowed. In terms
of quantum mechanical few- and many-body states this should be directly addressible 
using the techniques discussed above. An interesting recent proposal 
concerns a generalization to a system where there are three helical traps
intertwined corresponding to the geometry of a three-stranded DNA molecules. 
In this case it has been suggested that three-body bound states across 
the strands could occur in the case where no two-body bound states
across any two out of the three strands are possible \cite{maji2010,pal2013}.
This could connect such systems to the Efimov physics studied in 
few-body nuclear and atomic physics \cite{zinner2013}. Again the 
methods mentioned above would be able to address such ideas.

The authors acknowledge several inspiring conversations with G. De Chiara and
support from the Danish Council for Independent Research.

\end{document}